\documentclass[twocolumn,prb,aps,superscriptaddress]{revtex4-1}
\usepackage[T1]{fontenc}
\usepackage[latin9]{inputenc}

\usepackage{amsmath}
\usepackage{amssymb}
\usepackage{bbm}
\usepackage{braket}
\usepackage{dsfont}
\allowdisplaybreaks

\usepackage{graphicx}
\usepackage[colorlinks=true,citecolor=blue,linkcolor=blue]{hyperref}

\newcommand{\equref}[1]{Eq.~(\ref{#1})}
\newcommand{\equsref}[2]{Eqs.~(\ref{#1}) and (\ref{#2})}
\newcommand{\secref}[1]{Sec.~\ref{#1}}
\newcommand{\figref}[1]{Fig.~\ref{#1}}

\newcommand{\refcite}[1]{Ref.~\onlinecite{#1}}
\newcommand{\refscite}[1]{Refs.~\onlinecite{#1}}

\newcommand{\appref}{Appendix}
\newcommand{\tableref}[1]{Table~\ref{#1}}

\newcommand{\pdagger}{{\phantom{\dagger}}}
\newcommand{\Vint}{V}

\newcommand{\sign}{\,\text{sign}}
\renewcommand{\approx}{\simeq}

\renewcommand{\vec}[1]{\boldsymbol{#1}}

\newcommand{\ie}{i.e.~}

\begin{document}
\title{Pair breaking in multi-orbital superconductors: an application to oxide interfaces}

\author{M.\,S.\ Scheurer}
\affiliation{Institut f\"ur Theorie der Kondensierten
Materie, Karlsruher Institut f\"ur Technologie, D-76131 Karlsruhe, Germany}

\author{M.\ Hoyer}
\affiliation{Institut f\"ur Theorie der Kondensierten
Materie, Karlsruher Institut f\"ur Technologie, D-76131 Karlsruhe, Germany}
\affiliation{Institut f\"ur Festk\"orperphysik, Karlsruher Institut f\"ur Technologie, D-76021 Karlsruhe, Germany}

\author{J.\ Schmalian}
\affiliation{Institut f\"ur Theorie der Kondensierten
Materie, Karlsruher Institut f\"ur Technologie, D-76131 Karlsruhe, Germany}
\affiliation{Institut f\"ur Festk\"orperphysik, Karlsruher Institut f\"ur Technologie, D-76021 Karlsruhe, Germany}

\date{\today}

\begin{abstract}
We investigate the impact of impurity scattering on superconductivity in an anisotropic multi-orbital model with spin-orbit coupling which describes the electron fluid at two-dimensional oxide interfaces. 
As the pairing mechanism is under debate, both conventional and unconventional superconducting states are analyzed. 
We consider magnetic and nonmagnetic spin-dependent intra- and interorbital scattering and discuss possible microscopic realizations leading to these processes.
It is found that, for magnetic disorder, the unconventional superconductor is protected against interband scattering and, thus, more robust than the conventional condensate. In case of nonmagnetic impurities, the conventional superconductor is protected as expected from the Anderson theorem and the critical scattering rate of the unconventional state is enhanced by a factor of four due to the spin-orbit coupling and anisotropic masses in oxide interfaces.
\end{abstract}
\maketitle

\section{Introduction}
To understand the properties of many of the recently discovered superconductors\cite{Hideo,Reyren}, a multi-band description, which is often tied to the existence of multiple orbital degrees of freedom, is essential. Taking into account several Fermi surfaces does not only yield quantitative corrections, but also makes qualitatively different concepts possible such as pairing states that have no analogue in a single-band picture. 
One famous example is the $s^{+-}$ state that has been proposed, e.g., in the iron-based superconductors\cite{spmprop1,spmprop2} and earlier in other systems\cite{GorkovExoticSC}. While the pairing field is constant on a given Fermi surface, it changes sign between the electron and hole band.  

The identification of a pairing state and mechanism constitutes a challenging task, in particular, in a multi-band system. The behavior of superconductivity under the influence of different types of disorder can give important hints about the order parameter and pairing glue. In case of conventional $s$-wave superconductivity, the transition temperature is only very weakly affected by the presence of nonmagnetic impurities except for very strong disorder where the mean-free path $l$ is comparable to or smaller than the Fermi wavelength $k_F^{-1}$. This is usually referred to as the Anderson theorem\cite{AT,ATAG1,ATAG2}. 
When magnetic impurities are considered, the transition temperature is suppressed\cite{AGLaw} and vanishes when $l$ becomes of the order of the superconducting coherence length $\xi$ even when $k_F l \gg 1$. Since various unconventional pairing states have been shown\cite{pwavescattering,Annettpenetration,MazinUnconv,SrRuO,Balatskyreview,HirschfeldReview} to be strongly suppressed already by nonmagnetic impurities, the sensitivity of the transition temperature to nonmagnetic disorder is commonly used as an indication of unconventional pairing. Unconventional superconductivity is thus expected to occur only in very clean systems. 	 

However, several studies\cite{MazinUnconv,SpinOrbitLocking,PnictidesScatteringDolgov,PnictidesScattering} of both magnetic and nonmagnetic disorder have revealed that, in some cases, unconventional superconductors can be much more robust than naively expected or even enjoy an analogue of the Anderson theorem. 
Refs.~\onlinecite{MazinUnconv} and \onlinecite{SpinOrbitLocking,PnictidesScatteringDolgov,PnictidesScattering} indicate that it is essential to take into account all characteristic details of the system, such as multiple orbitals and spin-orbit coupling, in order to judge reliably the robustness of unconventional superconductors in the presence of disorder. From a theoretical point of view, it then becomes more difficult to find an approach that captures the essential features but leads to results that are universal in the sense that they do not depend on irrelevant microscopic details.

One example of a recently discovered superconducting\cite{Reyren} system where the multi-orbital character\cite{Joshua2012,TrisconeConfExp}, spin-orbit coupling\cite{CavigliaSOC,ShalomSOC} as well as strongly anisotropic effective masses\cite{Santander-SyroAniso} seem to be crucial for understanding its properties is provided by the two-dimensional (2D) electron liquid that forms\cite{Ohtomo} at the interface between the insulators LaAlO$_2$ (LAO) and SrTiO$_2$ (STO). 
In \refcite{ScheurerSchmalian}, two possible pairing states have been proposed for this system. If superconductivity arises as a consequence of conventional electron-phonon coupling, the pairing field will have the same sign on both spin-orbit split Fermi surfaces. In case of an unconventional, i.e., purely electronic, pairing mechanism, the order parameter has opposite signs on the two Fermi surfaces. Despite the similarities to the $s^{+-}$ state known from the iron-based superconductors, there is one main crucial difference. The Fermi surfaces in LAO/STO interfaces are singly degenerate due to the combination of broken inversion symmetry and spin-orbit coupling which has important consequences for superconductivity (see also, e.g., \refscite{GorkovRashba,DesignPrinciples}).
For example, the unconventional pairing state proposed\cite{ScheurerSchmalian} for oxide interfaces has been shown to be a time-reversal invariant topological superconductor which is intimately related to the Fermi surfaces being split by spin-orbit coupling.    

Consequently, it constitutes an important open question whether such an exciting new state of matter is indeed realized in oxide interfaces. 
An analysis of the impact of disorder on the superconducting states proposed in \refcite{ScheurerSchmalian} is thus required. It is particularly important to perform a quantitative analysis beyond a simple one-orbital toy model calculation since experimental estimates of the coherence length\cite{Reyren} $\xi \approx 70 - 105\, \textrm{nm}$ and the mean-free path\cite{ShalomMFP,ChangMFP} $l \approx 25 - 95\, \textrm{nm}$ are of the same order. 

In this paper, we consider impurity scattering in superconducting LAO/STO interfaces using a two-band model that captures the multi-orbital character, the spin-orbit splitting as well as the anisotropy of the masses.
The resulting highly non-circular form of the Fermi surfaces and anisotropic textures of the wavefunctions are conveniently taken into account by introducing patches in the nested regions of the Fermi surfaces. Due to the strong anisotropy, the wavefunctions are approximately constant within each patch yielding results that only depend on a small set of parameters. We find that the spin and orbital polarization of the states at the Fermi surfaces enhances the stability of both the conventional and unconventional superconducting phase against impurity scattering.
Furthermore, contrary to common wisdom, the unconventional superconductor is found to be, by a factor of two, more robust against magnetic scattering than the conventional condensate.

The remainder of the paper is organized as follows. In \secref{CleanSys}, we introduce the clean model of superconductivity. The impurity configurations consistent with the symmetries of the system and possible microscopic realizations thereof are discussed in \secref{DisorderConfigs}. In \secref{TechnicalPart}, we present the approximations used to calculate the transition temperature in the disordered system. Finally, the results can be found in \secref{Results}.

\section{Superconductivity in the clean system}
\label{CleanSys}
To describe superconductivity in oxide heterostructures, we use an effective Hamiltonian of the Ti $3d_{xz}$ and $3d_{yz}$ orbitals.
To justify why these are the essential degrees of freedom, let us recall that the orbitals relevant for the electronic low-energy physics are the Ti $3d$ states of the $t_{2g}$ manifold, \ie the $3d_{xy}$, $3d_{xz}$ and $3d_{yz}$ orbitals.
The states of the interface electron system derived from the $3d_{xy}$ orbital mainly localize and possibly order magnetically\cite{PentchevaPickett1,PentchevaPickett2,PotterLee,SpinSplitting}. This explains the discrepancy between the carrier concentration in the two-dimensional electron liquid expected from the polar catastrophe mechanism and that seen in transport measurements. The electronic states that lead to superconductivity are therefore most likely due to the $3d_{xz}$ and $3d_{yz}$ orbitals which is supported by recent experiments\cite{Joshua2012,TrisconeConfExp}. 

Let us first focus on the non-interacting part of the effective Hamiltonian of these states,
\begin{equation}
 H_0 = \sum_{\vec{k}}\psi_{\alpha,\vec{k}}^\dagger h_{\alpha\alpha'}(\vec{k}) \psi_{\alpha',\vec{k}}^\pdagger, \label{NormalStateHam}
\end{equation} 
where $\psi_{\alpha,\vec{k}}$ and $\psi^\dagger_{\alpha,\vec{k}}$ describe the annihilation and creation of quasi-particles of crystal momentum $\vec{k}$ in state~$\alpha$, with $\alpha$ labeling the four distinct combinations of the two orbitals ($3d_{xz}$, $3d_{yz}$) and the spin orientations.
The Hamiltonian is characterized by two ingredients, $h(\vec{k}) = h_{\text{m}}(\vec{k}) + h_{\text{so}}(\vec{k})$. 

Firstly, the different overlap of the two orbitals along the $x$- and $y$-direction leads to anisotropic masses described by 
\begin{equation}
 h_{\text{m}}(\vec{k}) = \begin{pmatrix} \frac{k_1^2}{2m_{\text{l}}} + \frac{k_2^2}{2m_{\text{h}}} & \delta k_1 k_2 \\ \delta k_1 k_2 & \frac{k_1^2}{2m_{\text{h}}} + \frac{k_2^2}{2m_{\text{l}}} \end{pmatrix} \otimes \sigma_0,
\end{equation} 
where the upper (lower) component refers to the $3d_{xz}$ ($3d_{yz}$) orbital and $\sigma_0$ is the unit matrix in spin space. Experimentally, large anisotropy ratios $m_{\text{h}}/m_{\text{l}} \approx 15 \dots 30$ have been reported\cite{Santander-SyroAniso}. The orbital mixing term $\delta$ describes interorbital second-nearest neighbor hopping. 

Secondly, the properties of the electron liquid are strongly affected by spin-orbit coupling \cite{CavigliaSOC,ShalomSOC}. As shown in \refcite{ScheurerSchmalian}, this leads to the contribution
\begin{align}
\begin{split}
h_{\text{so}}(\vec{k}) & =\frac{1}{2}\beta\tau_{2}\sigma_{3}+\alpha_{1}\tau_{0}\left(k_{1}\sigma_{2}-k_{2}\sigma_{1}\right)\\
 & +\alpha_{2}\tau_{1}\left(k_{1}\sigma_{1}-k_{2}\sigma_{2}\right)+\alpha_{3}\tau_{3}\left(k_{1}\sigma_{2}+k_{2}\sigma_{1}\right)\label{eq:spinorbfin}
\end{split}
\end{align}
to the Hamiltonian. Here $\alpha_i$ and $\beta$ are real constants and $\sigma_i$ ($\tau_i$) denote Pauli matrices in spin (orbital) space. We see that besides the usual Rashba coupling ($\propto \alpha_{1}$), the multi-orbital character of the model also allows for the atomic spin-orbit term proportional to $\beta$ and two Dresselhaus couplings ($\alpha_2$, $\alpha_3$). \equref{eq:spinorbfin} is the most general spin-orbit coupling Hamiltonian (up to linear order in $\vec{k}$) consistent with time-reversal and the $C_{4v}$ point symmetry of the system. Starting from a three-orbital model that also includes the $3d_{xy}$ band and integrating out\cite{ScheurerSchmalian} the latter reproduces \equref{eq:spinorbfin} with the additional constraint $\alpha_1=-\alpha_2=-\alpha_3$.

Note that we do not include any coupling to the magnetic moments that possibly develop in the $3d_{xy}$ bands. The reason is that no significant correlation between the inhomogeneous magnetization of the $3d_{xy}$ orbitals and the superconducting response has been observed\cite{BertPatches} and, hence, the effective exchange coupling between the localized magnetic states and the superconducting electrons is expected to be negligible for describing superconductivity. This can be understood as a consequence of the spatial separation of the localized magnetic and the itinerant superconducting bands and the small orbital admixture of $3d_{xy}$ in the states that host superconductivity. 

\begin{figure}[tb]
\begin{center}
\includegraphics[width=0.80\linewidth]{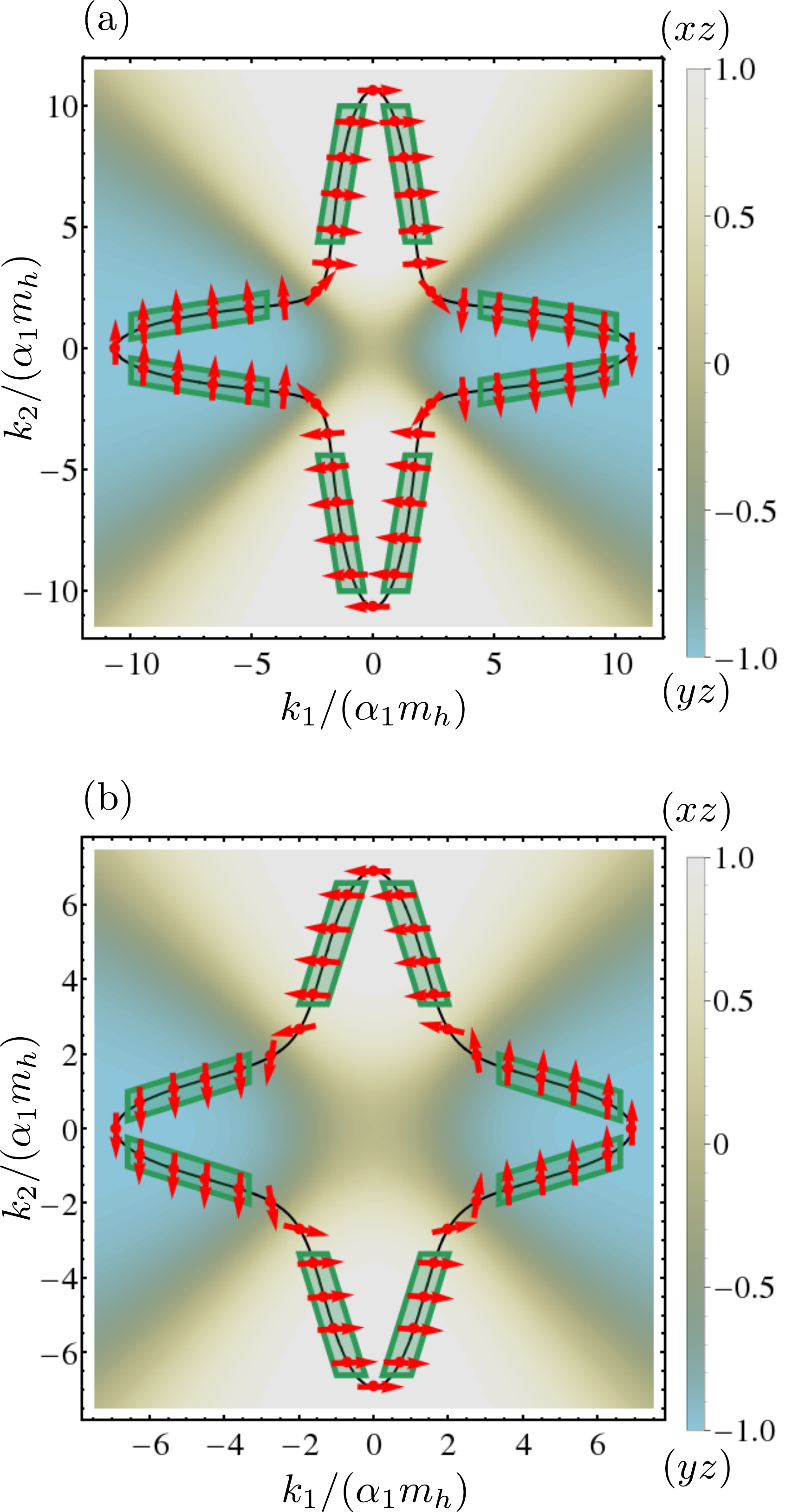}
\caption{Outer (a) and inner (b) Fermi surface (black lines) with the corresponding spin (red arrows) and orbital (color) polarizations within the two-orbital model~(\ref{NormalStateHam}). The patches with approximately constant wavefunctions are indicated in green. Here we have used $\beta=20\,\textrm{meV}$, $\alpha_1 = 10\,\textrm{meV\AA{}}$, $m_{\text{h}}/m_{\text{l}} = 30$, $m_{\text{l}}=0.7 m_{e}$ ($m_{e}$ is the free electron mass) and $\delta = 0.28 m_{e}^{-1}$ deduced from \refscite{Zhong,CavigliaSOC,Santander-SyroAniso,Joshua2012}.}
\label{SpinTexturesOfTheModel}
\end{center}
\end{figure}

The spin-orbit coupling (\ref{eq:spinorbfin}) breaks inversion symmetry and removes the spin-degeneracy of the Fermi surfaces. For chemical potentials close (deviation $<\beta$) to the bottom of the spectrum of $h$, only two out of the four singly degenerate bands of \equref{NormalStateHam} are relevant. As carrier concentration dependent studies\cite{Joshua2012,TrisconeConfExp} show that the sweet spot of superconductivity is associated with the Fermi level entering the $xz$- and $yz$-bands, we will focus on the two bands of $h(\vec{k})$ with lower energy. The corresponding form, spin and orbital polarization of the Fermi surfaces are shown in \figref{SpinTexturesOfTheModel}.  
In the $16$ patches, highlighted in green, the Fermi surfaces are nearly straight lines which are, in groups of four, approximately parallel (nesting).   
Within each patch, the wavefunctions are almost constant. The large mass anisotropy in $h_{\text{m}}$ leads to strong orbital polarization: The patches in the lobes along $k_1$ ($k_2$) are mainly $yz$-polarized ($xz$-polarized). Similarly, the spin-orbit coupling $h_{\text{so}}$ largely aligns the spin parallel or antiparallel to one of the in-plane coordinate axes. The strong spin and orbital 
polarization of the patch wavefunctions is of crucial relevance for impurity scattering as it can lead to the suppression of certain processes. This will be discussed in detail in \secref{Results}.

In \refcite{ScheurerSchmalian}, it has been shown that the system can have two distinct superconducting instabilities: In case of a conventional, electron-phonon induced, pairing mechanism, the pairing field has the same sign on both Fermi surfaces, which will be referred to as $s^{++}$ in the following. If the superconductor is due to the exchange of particle-hole fluctuations, the sign of the order parameter will change between the Fermi surfaces. Correspondingly, this unconventional superconducting state will be denoted by $s^{+-}$.   

At low energies, the relevant interaction channel that distinguishes between these two superconductors is pair hopping, i.e., the annihilation of a Kramers pair of electrons from one band and the creation of a Kramers pair in the other. Depending on the sign of the associated coupling constant, either the $s^{++}$ or $s^{+-}$ superconductor will be favored. As the sign conventions for this interaction term sensitively depends on the phase convention for the eigenstates of $h$, the mathematical formulation of the pair hopping interaction will be postponed to \secref{PhaseConvention}.

\section{Disorder configurations}
\label{DisorderConfigs}
Here we discuss the possible impurity configurations that can be present in oxide interfaces. The most general (quadratic) Hamiltonian of a given disorder realization~$W$ reads
\begin{equation}
 H_{\text{dis}} = \int_{\vec{x},\vec{x}'} \psi_{\alpha}^\dagger(\vec{x}) W_{\alpha\alpha'}(\vec{x},\vec{x}') \psi_{\alpha'}^\pdagger(\vec{x}'), \label{DisorderPotential}
\end{equation} 
where, for convenience, we have switched to a real-space representation with $\psi_{\alpha}(\vec{x})$ and $\psi_{\alpha}^\dagger(\vec{x})$ denoting the Fourier-conjugates of $\psi_{\alpha,\vec{k}}$ and $\psi^\dagger_{\alpha,\vec{k}}$, respectively. 

In what follows, we not only allow for disorder configurations that are consistent with the point group of the system but also take into account impurities that locally break any subset of the point symmetries. We only require these symmetries to be preserved on average. For example, oxygen vacancies that locally break the fourfold rotation symmetry occur with equal probability on bonds along the two perpendicular directions parallel to the interface. This effect is most easily discussed on the level of correlation functions. Focusing on Gaussian-distributed disorder, the entire information about the statistics is contained in the correlator
 \begin{align}
\begin{split}
 \Gamma&_{\alpha_1 \alpha_1',\alpha_2\alpha_2'}(\vec{x}_1,\vec{x}_1',\vec{x}_2,\vec{x}_2') \\ 
& \quad := \braket{W_{\alpha_1\alpha_1'}(\vec{x}_1,\vec{x}_1')W_{\alpha_2\alpha_2'}(\vec{x}_2,\vec{x}_2')}_\text{dis}.  \label{DefinitionCorrelator} \end{split}
\end{align}
Here $\braket{\dots}_\text{dis}$ represents the disorder average only taking into account Hermitian disorder configurations, $W = W^\dagger$. The latter constraint makes the averaging procedure equivalent to a Gaussian integration over a real-valued field.

Within the replica approach\cite{Replica}, averaging over $W$ leads to an effective impurity-induced electron-electron interaction that reads in the action description as 
\begin{align}
\begin{split}
 &S_{\text{R}}[\psi,\overline{\psi}]= -\frac{1}{2} \sum_{r,r'=1}^R \int_{\tau,\tau'}\int_{\vec{x}_1,\vec{x}_1',\vec{x}_2,\vec{x}_2'}  \\ 
& \qquad \times \overline{\psi}_{\alpha_1}^r(\vec{x}_1,\tau) \psi_{\alpha_1'}^r(\vec{x}_1',\tau) \overline{\psi}_{\alpha_2}^{r'}(\vec{x}_2,\tau') \psi_{\alpha_2'}^{r'}(\vec{x}_2',\tau')  \\ & \qquad \times \Gamma_{\alpha_1 \alpha_1',\alpha_2\alpha_2'}(\vec{x}_1,\vec{x}_1',\vec{x}_2,\vec{x}_2'). \,  \label{GeneralFormOfIneraction}\end{split}
\end{align}
Here, $\psi_\alpha^r$, $\overline{\psi}_\alpha^r$ are the Grassmann analogues of $\psi_{\alpha}$, $\psi_{\alpha}^\dagger$ with the additional replica index $r=1,\dots,R$, where $R$ denotes the number of replicas. At the end of the calculation of physical quantities, the limit $R \rightarrow 0$ has to be taken.
We see that the correlator defined in \equref{DefinitionCorrelator} plays the role of the bare disorder vertex function. The same holds in the diagrammatic disorder-averaging technique of \refcite{ADGBook}.

For simplicity, we will restrict our analysis to spatially local disorder, \ie $\vec{x}_j=\vec{x}'_j$ in \equref{GeneralFormOfIneraction}, and $\delta$-correlated statistics which, in addition, forces $\vec{x}_1 = \vec{x}_2$. Furthermore, let us assume the disorder distribution to be homogeneous. Taken together, the impurity vertex becomes spatially local and translation invariant,
\begin{align}
\begin{split}
 \Gamma&_{\alpha_1 \alpha_1',\alpha_2\alpha_2'}(\vec{x}_1,\vec{x}_1',\vec{x}_2,\vec{x}_2') \\ 
&= \delta(\vec{x}_1-\vec{x}_1')\delta(\vec{x}_2-\vec{x}_2')\delta(\vec{x}_1-\vec{x}_2) \Gamma'_{\alpha_1 \alpha_1',\alpha_2\alpha_2'}. \label{GeneralApprox}\end{split}
\end{align}     

\subsection{Symmetry properties of the impurity vertex}
To derive and classify the disorder configurations relevant for oxide interfaces, we have to analyze how the vertex $\Gamma'_{\alpha_1 \alpha_1',\alpha_2\alpha_2'}$ transforms under the point group operations and time-reversal. 

To begin with the former, consider an arbitrary operation $g$ of the \textit{point group}. The field operators transform according to
\begin{subequations}
\begin{align}
 \psi_\alpha(\vec{x}) \, &\stackrel{g}{\rightarrow} \, \left(\mathcal{R}^\dagger_\psi(g)\right)_{\alpha\beta} \psi_\beta(\mathcal{R}_v(g)\vec{x}), \\
 \psi^\dagger_\alpha(\vec{x}) \, &\stackrel{g}{\rightarrow} \, \psi^\dagger_\beta(\mathcal{R}_v(g)\vec{x}) \left(\mathcal{R}_\psi(g)\right)_{\beta\alpha},\end{align}\label{pointSymRepPath}\end{subequations}
where $\mathcal{R}_\psi^\dagger(g) \vec{J} \mathcal{R}_\psi(g) = \mathcal{R}_s(g) \vec{J}$ with $\mathcal{R}_v(g)$ and $\mathcal{R}_s(g)$ denoting the representation of $g$ on vectors and pseudovectors, respectively. Here $\vec{J}$ represents the total angular momentum operator.
From \equref{DisorderPotential} readily follows that
\begin{equation}
 W(\vec{x},\vec{x}') \, \stackrel{g}{\rightarrow} \, \mathcal{R}_\psi(g) W\left(\mathcal{R}^{-1}_v(g)\vec{x},\mathcal{R}^{-1}_v(g)\vec{x}'\right) \mathcal{R}_\psi^\dagger(g),
\end{equation}
which immediately yields the transformation behavior of the correlator (\ref{DefinitionCorrelator}).
As mentioned, while a generic given disorder realization $W$ will break all point symmetries of the system, on average the full point symmetry of the normal state has to be restored. In other words, the correlator (\ref{DefinitionCorrelator}) must transform trivially which leads, within the approximation (\ref{GeneralApprox}), to the constraint
\begin{align}
\begin{split}
 &\Gamma'_{\alpha_1 \alpha_1',\alpha_2\alpha_2'} \stackrel{!}{=} \left(\mathcal{R}_\psi(g)\right)_{\alpha_1\widetilde{\alpha}_1} \left(\mathcal{R}_\psi(g)\right)_{\alpha_2\widetilde{\alpha}_2} \\ & \qquad \quad \times \Gamma'_{\widetilde{\alpha}_1 \widetilde{\alpha}_1',\widetilde{\alpha}_2\widetilde{\alpha}_2'} \left(\mathcal{R}_\psi^\dagger(g)\right)_{\widetilde{\alpha}_2'\alpha_2'} \left(\mathcal{R}_\psi^\dagger(g)\right)_{\widetilde{\alpha}_1'\alpha_1'} \label{PointSymmetriesConstr}\end{split}
\end{align} 
for all operations $g$.

To distinguish between nonmagnetic and magnetic disorder, we have to take into account the properties under \textit{time reversal}. Let us denote the representation of time reversal on wavefunctions by $\Theta = \mathcal{T}\mathcal{K}$ with unitary $\mathcal{T}$ and $\mathcal{K}$ representing complex conjugation. Within the basis used in \equref{eq:spinorbfin}, it holds $\mathcal{T}=i\tau_0\sigma_2$. According to \equref{DisorderPotential}, the disorder Hamiltonian transforms under time-reversal as
\begin{equation}
 W(\vec{x},\vec{x}') \, \stackrel{\Theta}{\rightarrow} \, \mathcal{T} W^T(\vec{x}',\vec{x}) \mathcal{T}^{-1}.
\end{equation} 
Here we have already taken advantage of the Hermiticity of $W$ which allows for a representation that does not explicitly involve complex conjugation. This is convenient as the restriction on time-reversal symmetric (TRS), $\sigma = 1$, or time-reversal antisymmetric (TRA), $\sigma=-1$, disorder realizations can then be reformulated in terms of the constraints 
\begin{align}
\begin{split}
 \Gamma'_{\alpha_1 \alpha_1',\alpha_2\alpha_2'} &\stackrel{!}{=} \sigma \, \mathcal{T}_{\alpha_1\tilde{\alpha}_1} \Gamma'_{\tilde{\alpha}_1' \tilde{\alpha}_1,\alpha_2\alpha_2'} \left(\mathcal{T}^{-1}\right)_{\tilde{\alpha}_1'\alpha_1'}   \\ &\stackrel{!}{=} \sigma \, \mathcal{T}_{\alpha_2\tilde{\alpha}_2} \Gamma'_{\alpha_1\alpha_1',\tilde{\alpha}_2' \tilde{\alpha}_2} \left(\mathcal{T}^{-1}\right)_{\tilde{\alpha}_2'\alpha_2'} \label{TRBehviorOfCorr}\end{split}
\end{align}
on the disorder vertex. Note that it is sufficient to impose the condition in the first line as the second line automatically follows from the property
\begin{equation}
 \Gamma'_{\alpha_1 \alpha_1',\alpha_2\alpha_2'} = \Gamma'_{\alpha_2 \alpha_2',\alpha_1\alpha_1'}, \label{GenPropOfGamma}
\end{equation}   
which is a direct consequence of the definition (\ref{DefinitionCorrelator}).

Because of the very different behavior of superconductivity in the two cases, we will discuss TRS and TRA disorder separately, which will be referred to as ``nonmagnetic'' and ``magnetic'' in the following.

\subsection{Nonmagnetic disorder}
\label{NonMagDisorder}
To write down the most general disorder vertex $\Gamma'_{\alpha_1 \alpha_1',\alpha_2\alpha_2'}$ satisfying \equsref{PointSymmetriesConstr}{TRBehviorOfCorr} with $\sigma=1$, we perform an expansion in terms of Hermitian matrices $\{w_\mu\}$, i.e., let
\begin{equation}
 \Gamma'_{\alpha_1 \alpha_1',\alpha_2\alpha_2'} = \sum_{\mu,\mu'} C_{\mu \mu'} (w_{\mu})_{\alpha_1\alpha_1'} (w_{\mu'})_{\alpha_2\alpha_2'}. \label{GammaExpansionRed}
\end{equation}
Note that $\Gamma'_{\alpha_1 \alpha_1',\alpha_2\alpha_2'} = {\Gamma'}^*_{\alpha_1' \alpha_1,\alpha_2'\alpha_2}$ as a consequence of $W^\dagger = W$ in \equref{DefinitionCorrelator}. Together with \equref{GenPropOfGamma} this forces $C$ to be real and symmetric,
\begin{equation}
 C = C^* = C^T. \label{ConstraintOnC}
\end{equation} 
\begin{table}[tb]
\begin{center}
\caption{The symmetry properties of the basis functions constructed from the Pauli matrices in spin ($\sigma_i$) and orbital ($\tau_i$) space. Here $A_1$, $A_2$, $B_1$, $B_2$ and $E$ denote the irreducible representations of $C_{4v}$, whereas TRS (TRA) indicates that the matrix is even (odd) under time-reversal.}
\label{BasisFunctionsTrafo}
 \begin{tabular}{c|ccc} \hline \hline
    $\otimes$   & $\sigma_0$ & $\sigma_1,\sigma_2$ & $\sigma_3$ \\ \hline
$\tau_0$    & $A_1$/TRS & $E$/TRA     &   $A_2$/TRA       \\
$\tau_1$    & $B_2$/TRS   & $E$/TRA    & $B_1$/TRA    \\
$\tau_2$    & $A_2$/TRA   & $E$/TRS   & $A_1$/TRS  \\ 
$\tau_3$    & $B_1$/TRS   & $E$/TRA   & $B_2$/TRA  \\ \hline \hline
 \end{tabular}
\end{center}
\end{table}

Let us represent the basis matrices $\{w_\mu\}$ as the $16$ possible tensor products of the Pauli matrices $\sigma_i$ and $\tau_i$, $i=0,1,2,3$, referring to the spin and orbital degrees of freedom as in \equref{eq:spinorbfin}. 
This constitutes a convenient choice since these basis matrices both transform under the irreducible representations of the point group $C_{4v}$ and are either symmetric or antisymmetric under time-reversal as summarized in \tableref{BasisFunctionsTrafo}. Recalling \equref{ConstraintOnC}, one can then readily construct the most general vertex in case of nonmagnetic disorder,
\begin{align}
\begin{split}
 &\Gamma_{\alpha_1 \alpha_1',\alpha_2\alpha_2'}' \\ 
&= \gamma^{\text{I}}_{A_1} \left(\tau_0 \sigma_0\right)_{\alpha_1\alpha'_1} \left(\tau_0 \sigma_0\right)_{\alpha_2\alpha'_2} + \gamma^{\text{II}}_{A_1} \left(\tau_2 \sigma_3\right)_{\alpha_1\alpha'_1} \left(\tau_2 \sigma_3\right)_{\alpha_2\alpha'_2} \\
& + \frac{\gamma_{A_1}^{\text{III}}}{2} \left[\left(\tau_0 \sigma_0\right)_{\alpha_1\alpha'_1} \left(\tau_2 \sigma_3\right)_{\alpha_2\alpha'_2} + \left(\tau_2 \sigma_3\right)_{\alpha_1\alpha'_1} \left(\tau_0 \sigma_0\right)_{\alpha_2\alpha'_2}\right] \\
&+ \gamma_{B_1} \left(\tau_3 \sigma_0\right)_{\alpha_1\alpha'_1} \left(\tau_3 \sigma_0\right)_{\alpha_2\alpha'_2} + \gamma_{B_2} \left(\tau_1 \sigma_0\right)_{\alpha_1\alpha'_1} \left(\tau_1 \sigma_0\right)_{\alpha_2\alpha'_2} \\
&+ \frac{\gamma_E}{2} \left[ \left(\tau_2 \sigma_1\right)_{\alpha_1\alpha'_1} \left(\tau_2 \sigma_1\right)_{\alpha_2\alpha'_2} + \left(\tau_2 \sigma_2\right)_{\alpha_1\alpha'_1} \left(\tau_2 \sigma_2\right)_{\alpha_2\alpha'_2} \right]\end{split} \label{GeneralNonMagn}
\end{align}
characterized by six independent real coupling constants.
To gain physical insight into \equref{GeneralNonMagn}, let us first discuss the meaning of the different terms and then analyze how they can emerge microscopically.

The contributions proportional to $\gamma^{\text{I}}_{A_1}$, $\gamma_{B_1}$ and $\gamma_{B_2}$ correspond to spin-trivial disorder with orbital wavefunctions of different point symmetry: The different terms transform as $A_1$, $B_1$ and $B_2$ under $C_{4v}$ which corresponds to $s$-, $d_{x^2-y^2}$-, and $d_{xy}$-wave.
Noting that $\tau_2$ is the projection of the $z$-component of the orbital angular momentum operator on the subspace spanned by the $3d_{xz}$ and $3d_{yz}$ orbitals, the terms with $\gamma^{\text{II}}_{A_1}$ and $\gamma_E$ represent (atomic) spin-orbit disorder with out-of-plane/longitudinal and in-plane/transversal orientation of the spin, respectively.
The prefactors of the two terms in the last line of \equref{GeneralNonMagn} must be identical in order to ensure $C_4$-rotation symmetry of the disorder distribution.
Finally, the contribution proportional to $\gamma^{\text{III}}_{A_1}$ describes the fact that the $C_{4v}$ point group allows for an admixture of $s$-wave disorder and longitudinal spin-orbit disorder on the same site.  

\begin{figure}[tb]
\begin{center}
\includegraphics[width=\linewidth]{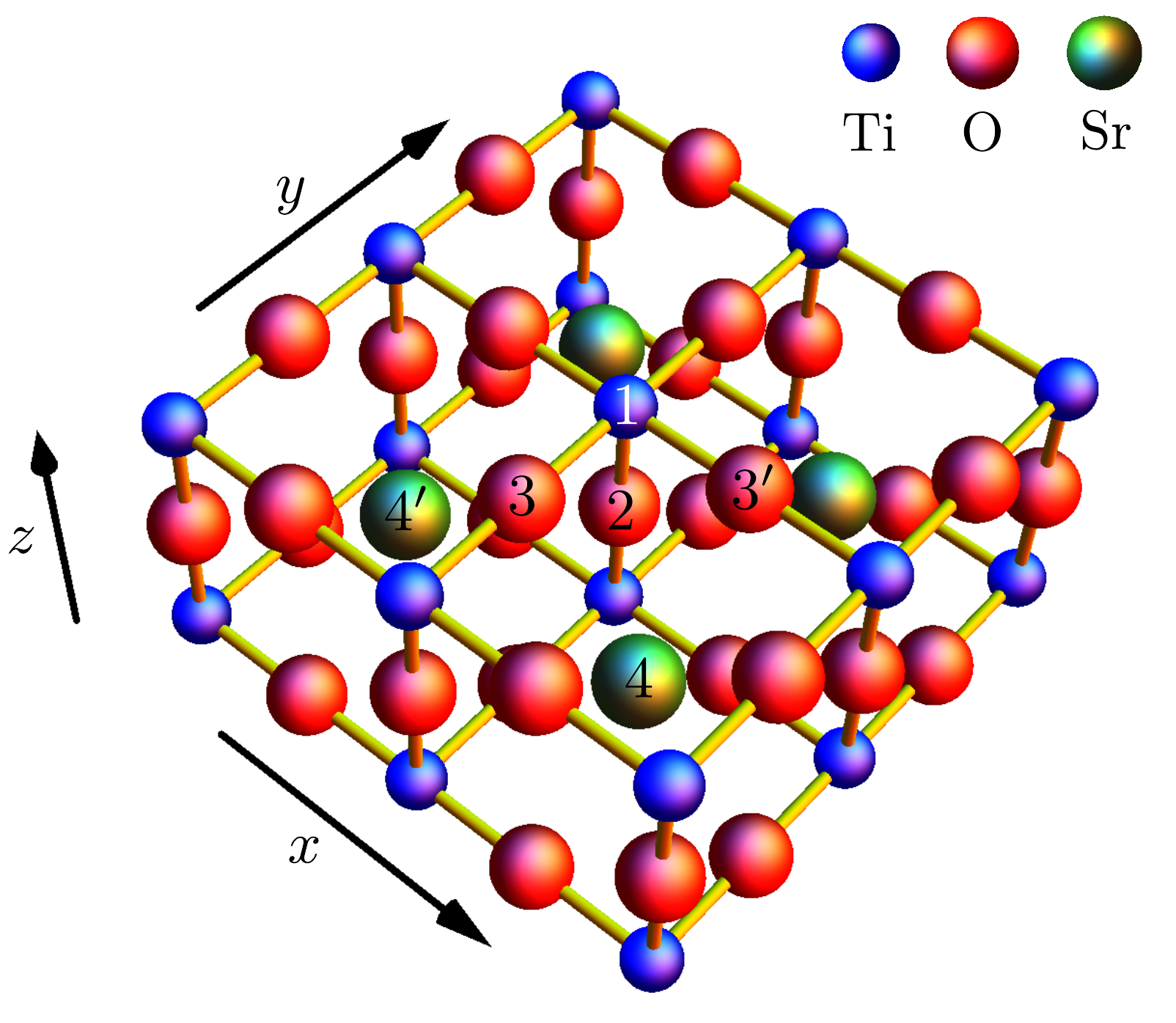}
\caption{Schematic of four unit cells of the crystal structure of STO. The part shown is assumed to be close to the interface (parallel to the $xy$-plane) in the spatial region where the conducting electron system resides. The numbers refer to the different positions of Ti ($1$), O ($2$, $3$, $3'$) and Sr ($4$, $4'$) which are referred to in the main text for the discussion of the different terms in the impurity vertices.}
\label{PossibleDefects}
\end{center}
\end{figure}

To understand how the different terms can be realized microscopically let us investigate the schematic illustration of the crystal structure in \figref{PossibleDefects}. We begin with oxygen vacancies, which are frequently discussed in the context of LAO/STO interfaces (see, e.g., \refscite{OxygenRijders,OxygenAriando}). Assume that one oxygen at position $2$, right below the central Ti at position $1$, is vacant. In this case, the defect preserves the full $C_{4v}$ point symmetry of the crystal such that only the first $3$ terms in \equref{GeneralNonMagn} can be present. The explicit values of $\gamma^{\text{I}}_{A_1}$, $\gamma^{\text{II}}_{A_1}$ and $\gamma^{\text{III}}_{A_1}$ depend on microscopic details of the system, however, as will be seen in \secref{Results}, only the sum of these scattering rates enters the effective pair-breaking strength. Here we mention that all coupling constants in the impurity vertex are proportional to the concentration of the defects considered. If instead an oxygen 
adjacent to the central Ti, e.g., at position $3$, is vacant, all point symmetries of the lattice are locally broken except for the mirror symmetry with respect to the $yz$-plane. This symmetry constraint rules out the contribution proportional to $\gamma_{B_2}$ and the first term in the last line of \equref{GeneralNonMagn} as both $\tau_1\sigma_0$ and $\tau_2\sigma_1$ are odd under $(x,y) \rightarrow (-x,y)$. Correspondingly, if we consider oxygen vacancies at position $3'$, the second term in the last line of \equref{GeneralNonMagn} and again the contribution with $\gamma_{B_2}$ will be forbidden. Recalling that the prefactors of the two terms in the last line of \equref{GeneralNonMagn} have to be equal by $C_4$ rotation symmetry, we see that, as expected, vacancies at position $3$ and $3'$ have to be included with equal probability to restore the full point symmetry after averaging. In summary, single oxygen vacancies allow for all contributions in \equref{GeneralNonMagn} except for the $\gamma_{B_2}$-
term.

To see how the latter can occur, let us now assume that a Sr atom is vacant, e.g., at position $4$ in \figref{PossibleDefects}. Similarly as above, all point symmetries are broken save for the mirror reflection at the $(110)$ plane. This now renders the $\gamma_{B_2}$-term possible which is even under the reflection at $(110)$. Assuming that, with the same probability, a Sr atom is missing at position $4'$, one restores rotation symmetry. In this case, all terms in \equref{GeneralNonMagn} are allowed except for the one proportional to $\gamma_{B_1}$. Finally, we note that the same is true for a local oxygen double vacancy where oxygen atoms are missing at position $3$ and $3'$ at the same time. We have thus provided examples of how all the terms in the most general non-magnetic impurity vertex (\ref{GeneralNonMagn}) can be realized.

\subsection{Magnetic disorder}
In case of magnetic disorder, \ie $\sigma=-1$ in \equref{TRBehviorOfCorr}, we proceed in the same way as in \secref{NonMagDisorder}.
From \tableref{BasisFunctionsTrafo}, one finds that there are three independent terms in the impurity vertex that correspond to in-plane magnetic moments,
\begin{subequations}
\begin{align}
\begin{split}
 &{\Gamma'}^\parallel_{\alpha_1 \alpha_1',\alpha_2\alpha_2'} \\ 
&= \frac{\gamma^{\text{I}}_{E}}{2}\left[ \left(\tau_0 \sigma_1\right)_{\alpha_1\alpha'_1} \left(\tau_0 \sigma_1\right)_{\alpha_2\alpha'_2} + \left(\tau_0 \sigma_2\right)_{\alpha_1\alpha'_1} \left(\tau_0 \sigma_2\right)_{\alpha_2\alpha'_2} \right] \\
& +\frac{\gamma^{\text{II}}_{E}}{2}\left[ \left(\tau_1 \sigma_1\right)_{\alpha_1\alpha'_1} \left(\tau_1 \sigma_1\right)_{\alpha_2\alpha'_2} + \left(\tau_1 \sigma_2\right)_{\alpha_1\alpha'_1} \left(\tau_1 \sigma_2\right)_{\alpha_2\alpha'_2} \right] \\
& +\frac{\gamma^{\text{III}}_{E}}{2}\left[ \left(\tau_3 \sigma_1\right)_{\alpha_1\alpha'_1} \left(\tau_3 \sigma_1\right)_{\alpha_2\alpha'_2} + \left(\tau_3 \sigma_2\right)_{\alpha_1\alpha'_1} \left(\tau_3 \sigma_2\right)_{\alpha_2\alpha'_2} \right],\end{split} \label{MagneticImpPara}
\end{align} 
and five terms,
\begin{align}
\begin{split}
 &{\Gamma'}^\perp_{\alpha_1 \alpha_1',\alpha_2\alpha_2'} \\ 
 &= \gamma^{\text{I}}_{A_2} \left(\tau_0 \sigma_3\right)_{\alpha_1\alpha'_1} \left(\tau_0 \sigma_3\right)_{\alpha_2\alpha'_2} + \gamma_{B_1} \left(\tau_1 \sigma_3\right)_{\alpha_1\alpha'_1} \left(\tau_1 \sigma_3\right)_{\alpha_2\alpha'_2}  \\
& + \gamma_{B_2} \left(\tau_3 \sigma_3\right)_{\alpha_1\alpha'_1} \left(\tau_3 \sigma_3\right)_{\alpha_2\alpha'_2} + \gamma^{\text{II}}_{A_2} \left(\tau_2 \sigma_0\right)_{\alpha_1\alpha'_1} \left(\tau_2 \sigma_0\right)_{\alpha_2\alpha'_2}  \\
 &+ \frac{\gamma^{\text{III}}_{A_2}}{2}\left[ \left(\tau_2 \sigma_0\right)_{\alpha_1\alpha'_1} \left(\tau_0 \sigma_3\right)_{\alpha_2\alpha'_2} + \left(\tau_0 \sigma_3\right)_{\alpha_1\alpha'_1} \left(\tau_2 \sigma_0\right)_{\alpha_2\alpha'_2} \right],\end{split} \label{MagneticImpPerp}
\end{align}\label{MagneticImp}\end{subequations}
where the magnetic moment is oriented perpendicular to the plane. As before, all coupling constants must be real due to \equref{ConstraintOnC}.

Physically, the three terms $\gamma^{\text{I}}_{E}$, $\gamma^{\text{II}}_{E}$ and $\gamma^{\text{III}}_{E}$ in ${\Gamma'}^\parallel$ correspond to impurities with in-plane spin-magnetization and distinct orbital symmetry ($s$-, $d_{xy}$- and $d_{x^2-y^2}$-wave). 
Note that no orbital-magnetic disorder with in-plane orientation can occur in the two-orbital model as the projection of the $x$- and $y$-component of the orbital angular momentum operator onto the $3d_{xz}$ and $3d_{yz}$ subspace vanishes identically.
The analogous terms in ${\Gamma'}^\perp$ with the same orbital symmetry but spin polarization along the $z$-direction are proportional to $\gamma^{\text{I}}_{A_2}$, $\gamma_{B_1}$ and $\gamma_{B_2}$, respectively. Finally, $\gamma^{\text{II}}_{A_2}$ corresponds to purely orbital magnetism along the $z$-direction and $\gamma^{\text{III}}_{A_2}$ shows that an arbitrary admixture of spin and orbital magnetism along the $z$-direction is allowed by symmetry.

Similarly to the nonmagnetic case, let us discuss some examples of how the terms in \equref{MagneticImp} can emerge microscopically. 
One can imagine that locally one Ti atom orders magnetically due to the proximity of the system to a competing magnetic instability\cite{ScheurerSchmalian}.
Numerical calculations indicate\cite{DFToxygen1,DFToxygen2} that magnetic moments on Ti atoms can be induced by the presence of oxygen vacancies.
Alternatively, we may think of Ti being replaced by a different magnetic atom.
In all cases, the form of the resulting impurity potential crucially depends on the orbitals that host the magnetic moment and whether it is orbital or spin magnetism. Thus, in general, all terms in \equref{MagneticImp} are feasible. To be more specific, recalling that the observed magnetism at the interface is mainly due to the $3d_{xy}$ band\cite{PentchevaPickett1,PentchevaPickett2,PotterLee,SpinSplitting}, one may expect that, most likely, the spin degree of freedom of the Ti $3d_{xy}$ orbital orders locally. In that case, the impurity vertex would be described by the terms proportional $\gamma_E^{\text{II}}$ and $\gamma_{B_1}$ as the associated orbital structure, $\tau_1$, transforms as $xy$ under $C_{4v}$.

If, instead, a magnetic moment develops at a site that does not lie at a high symmetry point with respect to the Ti atoms, the symmetry constraints will be less restrictive. Assuming, e.g., a spin-magnetic moment along the $z$-axis with $s$-wave orbital symmetry (\ie $\tau_0\sigma_3$) at the oxygen site $3$ ($3'$) in \figref{PossibleDefects}, the disorder potential is only restricted to be odd both under time-reversal and mirror reflection symmetry with respect to the $yz$-plane ($xz$-plane). The reduction of symmetry constraints simply follows from the fact that, irrespective of the symmetries of the local impurity, the presence of a perturbation at $3$ ($3'$) already breaks all symmetries except for time-reversal and one mirror reflection. Under these two remaining symmetry operations, the resulting impurity potential in the effective model for the conducting Ti orbitals must have the same behavior as $\tau_0\sigma_3$ (in both cases being odd). From these criteria one finds that 
only the term proportional to $\gamma_{B_1}$ can be excluded in \equref{MagneticImp}.

\section{Critical temperature of the disordered superconductor}
\label{TechnicalPart}
In this section, we provide details about how the critical temperature of the superconductors is calculated in the presence of disorder. 
Readers that are mainly interested in the results of this paper and the physical interpretation thereof can skip this part and directly proceed with \secref{Results}.

\subsection{Patch approximation}
\label{PatchApprox}
\begin{figure}[tb]
\begin{center}
\includegraphics[width=0.9\linewidth]{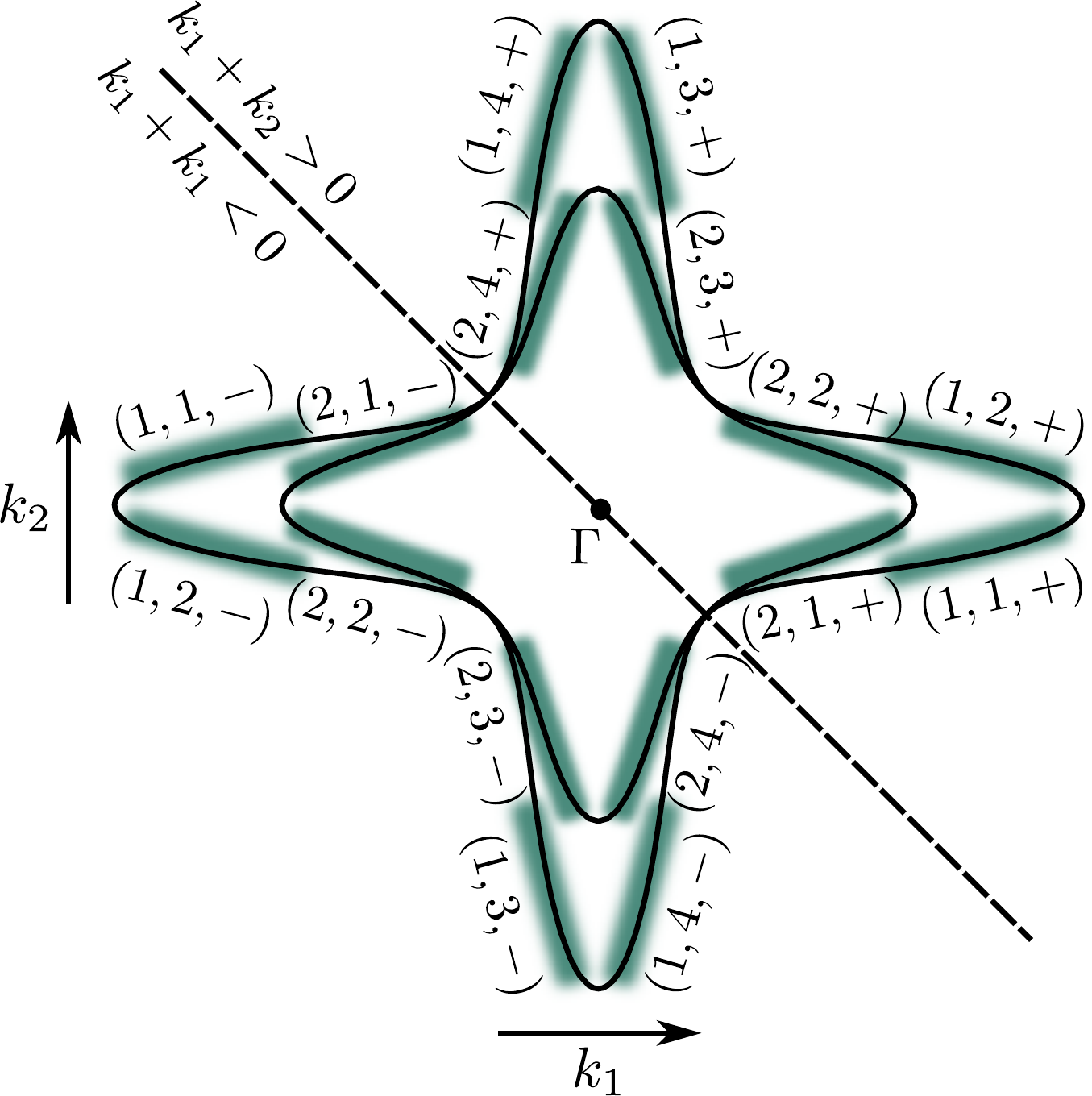}
\caption{Illustration of the different patches (green shaded domains) on the Fermi surfaces and the notation we use to label them. The patches are indicated by a composite index $\tau\equiv(\lambda,j,\eta)$ with $\lambda$, $j$, and $\eta$ referring to the two Fermi surfaces, the four patches with fixed sign of $k_1+k_2$, and to the two half-spaces, respectively.}
\label{IllustrationOfPatchNotation}
\end{center}
\end{figure}

For our calculation, it will be convenient to work in the eigenbasis of the normal state Hamiltonian $h(\vec{k})$ in \equref{NormalStateHam}. 
For this purpose, we introduce new operators $f_{\alpha,\vec{k}}$ and $f^\dagger_{\alpha,\vec{k}}$ via
\begin{equation}
 \psi_{\alpha,\vec{k}} = \left(\phi_{\beta,\vec{k}}\right)_\alpha f_{\beta,\vec{k}}, \quad \psi^\dagger_{\alpha,\vec{k}} = f^\dagger_{\beta,\vec{k}} \left(\phi^*_{\beta,\vec{k}}\right)_\alpha , \label{EigenbasisDef}
\end{equation} 
where $\phi_{\alpha,\vec{k}}$ are the four eigenstates of $h(\vec{k})$.
By design, the normal state Hamiltonian (\ref{NormalStateHam}) becomes diagonal,
\begin{equation}
 H_0 = \sum_{\vec{k}} \epsilon_\alpha(\vec{k}) f^\dagger_{\alpha,\vec{k}} f^\pdagger_{\alpha,\vec{k}} . \label{QuadraticHam}
\end{equation} 
In the new basis, the momentum space representation of the impurity vertex in \equref{GeneralApprox} reads
\begin{equation}
\left(\phi^*_{\alpha_1,\vec{k}_1}\right)_{\widetilde{\alpha}_1} \left(\phi^*_{\alpha_2,\vec{k}_2}\right)_{\widetilde{\alpha}_2} \Gamma'_{\widetilde{\alpha}_1 \widetilde{\alpha}_1',\widetilde{\alpha}_2 \widetilde{\alpha}_2'} \left(\phi_{\alpha_1',\vec{k}_1'}\right)_{\widetilde{\alpha}_1'} \left(\phi_{\alpha_2',\vec{k}_2'}\right)_{\widetilde{\alpha}_2'} 
\end{equation} 
which depends on all four external momenta. To simplify the analysis, we take advantage of the fact that, within each of the patches of the Fermi surfaces in \figref{SpinTexturesOfTheModel}, the wavefunctions $\phi_{\alpha,\vec{k}}$ are approximately constant. Together with the fact that this set of patches covers the main part of both Fermi surfaces, it is justified to replace  
\begin{align}
\begin{split}
 \phi_{\alpha,\vec{k}} \, \rightarrow \, \phi_{(\lambda,j,\eta)}, & \qquad f_{\alpha,\vec{k}} \, \rightarrow \, f_{(\lambda,j,\eta),\vec{p}} , \\ \sum_{\alpha,\vec{k}} \dots \, &\rightarrow \, \sum_{\lambda,j,\eta,\vec{p}} \dots\, .  \label{PatchApproxEq}\end{split}
\end{align} 
We only keep the fermions from the patches as dynamical degrees of freedom characterized by a constant spin and orbital polarization in a given path. This is what we refer to as the ``patch approximation''.
In \equref{PatchApproxEq} and in the following, $\lambda =1,2$ denotes the two Fermi surfaces and $j=1,2,3,4$, $\eta = +$ ($\eta=-$) represent the four distinct patches on each Fermi surface in the half-space with $k_1+k_2 > 0$ ($k_1+k_2 < 0$) as illustrated in \figref{IllustrationOfPatchNotation}. 
The additional index $\eta$ will be convenient for describing superconductivity as, for fixed $\lambda$ and $j$, the states with $\eta = +$ and $\eta=-$ are related by time reversal. We use $\vec{p}$ to denote the deviation of the crystal momentum from the centers of the different patches. We always assume that these momenta are cut off in a way that does not allow for patches to overlap.
Note that applying the patch approximation automatically takes advantage of the fact that, as discussed in \secref{CleanSys}, two out of the four bands of the normal state Hamiltonian (\ref{NormalStateHam}) can be neglected for energetic reasons.

Using the compact multi-index notation $\tau \equiv (\lambda,j,\eta)$, the disorder induced interaction (\ref{GeneralFormOfIneraction}) finally becomes within the patch approximation
\begin{align}
\begin{split}
 &S_{\text{R}}[\psi,\overline{\psi}] = -\frac{1}{2}  \sum_{r,r'=1}^R \int_{\tau,\tau'} \sum_{\vec{p}_1,\vec{p}_1',\vec{p}_2,\vec{p}_2'} \Gamma^{\text{PA}}_{\tau_1 \tau_1',\tau_2 \tau_2'} \\ 
&  \times \overline{f}_{\tau_1,\vec{p}_1}^r(\tau) f_{\tau_1',\vec{p}_1'}^r(\tau) \overline{f}_{\tau_2,\vec{p}_2}^{r'}(\tau') f_{\tau_2',\vec{p}_2'}^{r'}(\tau') \, D^{\tau_1\tau_1',\tau_2\tau_2'}_{\vec{p}_1\vec{p}_1',\vec{p}_2\vec{p}_2'}, \label{FinalDisorderInteraction}\end{split}
\end{align}
where, using the expansion (\ref{GammaExpansionRed}), 
\begin{align}
\begin{split}
\Gamma^{\text{PA}}_{\tau_1 \tau_1',\tau_2 \tau_2'} &=  \sum_{\mu,\mu'} C_{\mu \mu'} (\phi^\dagger_{\tau_1} w_\mu\phi_{\tau_1'}) (\phi^\dagger_{\tau_2}w_{\mu'}\phi_{\tau_2'})  \\
&= \,\, \vcenter{\hbox{\includegraphics[height=3.5em]{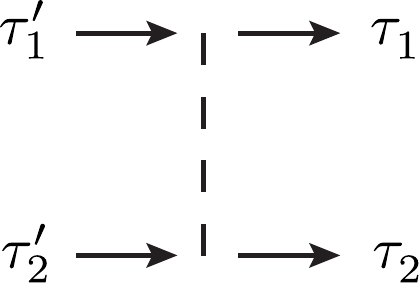}}} \label{VertexInEigenbasis}\end{split}
\end{align}
In the second line of \equref{VertexInEigenbasis}, we have introduced the diagrammatic symbol for the impurity vertex that will be used in \secref{GinzburgLandau}. 
Due to the translation invariance of the disorder distribution (see \equref{GeneralApprox}), the impurity induced interaction (\ref{FinalDisorderInteraction}) must conserve momentum, which is accounted for by $D^{\tau_1\tau_1',\tau_2\tau_2'}_{\vec{p}_1\vec{p}_1',\vec{p}_2\vec{p}_2'} \in \{0,1\}$. Here we will not specify the explicit form of $D$. Let us only note that it crucially depends on details of the band structure. To see this, consider, e.g., the scattering process of an electron from $(1,2,+)$ to $(2,2,+)$ with an electron that starts and does not leave the patch $(1,4,-)$ in \figref{IllustrationOfPatchNotation}. Obviously, the available phase space of this process depends both on the nesting vector between $(1,2,+)$ and $(2,2,+)$ as well as on the geometry of the patches. Fortunately, as we will explain in \secref{GinzburgLandau}, the transition temperature of the superconducting states will not depend on these details of the band structure.    

We emphasize that the vertex $\Gamma^{\text{PA}}$ contains the full information about disorder scattering within the patch approximation, combining knowledge about the microscopic structure (see \secref{DisorderConfigs}) of the impurities involved and the form of the eigenstates (see \figref{SpinTexturesOfTheModel}) between which electrons are scattered.
Motivated by the strong orbital and spin polarization of the patch wavefunctions, we approximate the states $\phi_{\tau}$ to be perfectly polarized. More precisely, within each patch, we take the spin to be fully aligned parallel to one of the coordinate axes and the orbital weight to be either fully $xz$ or $yz$ corresponding to the approximate polarization of the wavefunctions illustrated in \figref{SpinTexturesOfTheModel}. This simplification allows us to obtain analytic expressions for the scattering vertex~$\Gamma^{\text{PA}}$. In \secref{Results}, we will also comment on the impact of deviations from perfect polarization.

\subsection{Phase convention for the eigenstates}
\label{PhaseConvention}
To have a well-defined representation of the superconducting order parameter in the eigenbasis of the normal-state Hamiltonian, it is essential to constrain the phases of the eigenstates $\phi_{\alpha,\vec{k}}$ in \equref{EigenbasisDef}. To see this, consider the microscopic order parameter $\Delta_{\alpha\beta}(\vec{k})$ defined by the mean-field Hamiltonian   
\begin{equation}
 H_\Delta = \frac{1}{2} \sum_{\vec{k}} \left(\psi^\dagger_{\alpha,\vec{k}} \Delta_{\alpha\beta}(\vec{k})\psi^\dagger_{\beta,-\vec{k}} + \text{H.c.} \right), \label{OrderParamHam}
\end{equation}
which reads in the eigenbasis (\ref{EigenbasisDef}) as
\begin{equation}
 \widetilde{\Delta}_{\alpha,\vec{k}} := \phi^\dagger_{\alpha,\vec{k}} \Delta(\vec{k}) \phi^*_{\alpha,-\vec{k}}. \label{DeltaInEigenbasis}
\end{equation} 
Here it has already been taken into account that we focus on order parameters that have no Fermi-surface off-diagonal elements\cite{DesignPrinciples}. Clearly, \equref{DeltaInEigenbasis} is not invariant under a general $\vec{k}$-space local $U(1)$-gauge transformation $\phi_{\alpha,\vec{k}} \rightarrow e^{i \zeta^\alpha_{\vec{k}}} \phi_{\alpha,\vec{k}}$ of the wavefunctions. 

To remove this ambiguity, we take advantage of the fact that the normal state Hamiltonian $h(\vec{k})$ is even under time reversal. Since the Fermi surfaces are singly degenerate, one can always adjust the phases of the eigenstates to satisfy
\begin{equation}
  \Theta \phi_{(\lambda,j,\eta)} = -i\eta \phi_{(\lambda,j,-\eta)}, \label{PhaseConventionPatch}
\end{equation}
where, for convenience, we have switched to the patch notation. The phase factor $-i$ is actually arbitrary but has been chosen to reproduce the conventions of \refcite{ScheurerSchmalian}. Note that the sign factor $\eta$ is necessary due to the property $\Theta^2=-\mathds{1}$ of the time-reversal operator. Imposing \equref{PhaseConventionPatch} fixes the relative phase between $\phi_{\alpha,\vec{k}}$ and $\phi_{\alpha,-\vec{k}}$, thus, rendering \equref{DeltaInEigenbasis} gauge invariant.

To classify the possible superconducting phases, it is essential to know how the order parameter transforms under the elements of the point group of the system. Applying \equref{pointSymRepPath} in \equref{OrderParamHam} one readily finds that
\begin{equation}
 \Delta(\vec{k})  \, \stackrel{g}{\rightarrow} \,  \mathcal{R}_\psi(g) \Delta(\mathcal{R}_v^{-1}(g)\vec{k}) \mathcal{R}^T_\psi(g). \label{DeltaTrafoBeh}
\end{equation} 
Taking into account the phase convention (\ref{PhaseConventionPatch}) and using one more time that the Fermi surfaces are singly degenerate, it follows that the representation (\ref{DeltaInEigenbasis}) of the order parameter transforms according to (again switching to the patch notation)
\begin{equation}
 \widetilde{\Delta}_{(\lambda,j,\eta)} \, \stackrel{g}{\longrightarrow} \,  \eta \, \eta' \, \widetilde{\Delta}_{(\lambda',j',\eta')}, \label{TransformationPropsOfDelta}
\end{equation} 
where $(\lambda',j',\eta')$ denotes the patch obtained when applying the point group operation $g$ on the subset of momentum space belonging to patch $(\lambda,j,\eta)$.

General symmetry and energetic arguments\cite{DesignPrinciples} indicate that the superconducting order parameter of oxide interfaces transforms according to a one-dimensional irreducible representation of the point group $C_{4v}$. Experiments\cite{Richter2013} indicate that the superconductor is fully gapped. Together with the result that time-reversal symmetry is not broken\cite{DesignPrinciples}, the order parameter must transform under the trivial representation of $C_{4v}$ ($s$-wave) and, hence,
\begin{equation}
 \widetilde{\Delta}_{(\lambda,j,\eta)} = \eta \, \Delta_\lambda. \label{swavesymmetry}
\end{equation}
The microscopic calculation of \refcite{ScheurerSchmalian} yields $\Delta_1 = \Delta_2$ and $\Delta_1 = -\Delta_2$ corresponding to the $s^{++}$ and $s^{+-}$ superconductors as possible candidates. As already discussed in \secref{CleanSys}, the crucial interaction leading to these two instabilities is pair-hopping, which is given by
\begin{align}
\begin{split}
 H_{\text{int}} = \Vint \sum_{j,j'=1}^4 & \sum_{\vec{p},\vec{p}'} \Biggl(f^\dagger_{(1,j,+),\vec{p}}f^\dagger_{(1,j,-),-\vec{p}} \\ 
& \times f_{(2,j',-),-\vec{p}'}f_{(2,j',+),\vec{p}'} + \text{H.c.}\Biggr). \label{PairHoppingInteraction}
\end{split}
\end{align} 
Within the phase convention (\ref{PhaseConventionPatch}), taking $\Vint <0$ ($\Vint >0$) leads to the conventional $s^{++}$ (unconventional $s^{+-}$) superconductor.

\subsection{Ginzburg-Landau expansion in the presence of disorder}
\label{GinzburgLandau}
To determine the transition temperature of the superconductor, we first decouple the pair-hopping interaction~(\ref{PairHoppingInteraction}) by a mean-field approximation in the Cooper channel assuming $s$-wave symmetry as defined in \equref{swavesymmetry}. This yields
\begin{align}
 \begin{split}
H_{\text{int}}^{\text{MF}} = \frac{1}{2}&\sum_{\lambda,j,\eta}\sum_{\vec{p}} \Bigl(\eta \Delta_\lambda f^\dagger_{(\lambda,j,\eta),\vec{p}} f^\dagger_{(\lambda,j,-\eta),-\vec{p}}  \\ &+ \text{H.c.}\Bigr) - \frac{1}{\Vint } \left(\Delta_2^*\Delta_1^{\phantom{*}}+\Delta_1^* \Delta_2^{\phantom{*}}\right). \end{split}
\end{align}
Let us denote the free energy for a given disorder realization $W$ in \equref{DisorderPotential} by $\mathcal{F}[W,\Delta_\lambda]$. Assuming that the system is self-averaging, the free energy of the superconductor is given by the disorder average $\braket{\mathcal{F}[W,\Delta_\lambda]}_\text{dis}$ that only depends on the disorder correlator (\ref{DefinitionCorrelator}) and not on details of the microscopic $W$. As we are only interested in the transition temperature of superconductivity, it is sufficient to expand the free energy up to second order in the order parameters. One finds
\begin{equation}
 \braket{\mathcal{F}[W,\Delta_\lambda]}_\text{dis} \sim \Delta_{\lambda}^* d_{\lambda\lambda'}(T)  \Delta_{\lambda'}^{\phantom{*}} - \frac{1}{\Vint } \left(\Delta_2^*\Delta_1^{\phantom{*}}+\Delta_1^* \Delta_2^{\phantom{*}}\right), \label{AveragedFreeEnergy}
\end{equation}
as $\Delta_\lambda \rightarrow 0$, where the coefficients are given by
\begin{align}
 \begin{split}
&d_{\lambda\lambda'}(T) = -\frac{T}{4} \int_{\tau,\tau'} \Biggl\langle\Biggl\langle\sum_{j,\eta,\vec{p}}  \eta f_{(\lambda,j,-\eta),-\vec{p}}(\tau) f_{(\lambda,j,\eta),\vec{p}}(\tau)  \\ &  \times \sum_{j',\eta',\vec{p}'} \eta' \bar{f}_{(\lambda',j',\eta'),\vec{p}'}(\tau') \bar{f}_{(\lambda',j',-\eta'),-\vec{p}'}(\tau') \Biggr\rangle_{\hspace{-0.3em}0} \Biggr\rangle_{\hspace{-0.3em}\text{dis}} \end{split} \label{CoefficientsInExpansion}
\end{align} 
in the path integral representation. Here $\braket{\dots}_0$ denotes the expectation value with respect to the action associated with the normal state Hamiltonian (\ref{QuadraticHam}).
The transition temperature and the order parameter then simply follow from diagonalizing the quadratic form in \equref{AveragedFreeEnergy}. These two quantities are determined by the eigenvalue that first changes sign and the associated eigenvector, respectively.

\begin{figure}[tb]
\begin{center}
\includegraphics[width=\linewidth]{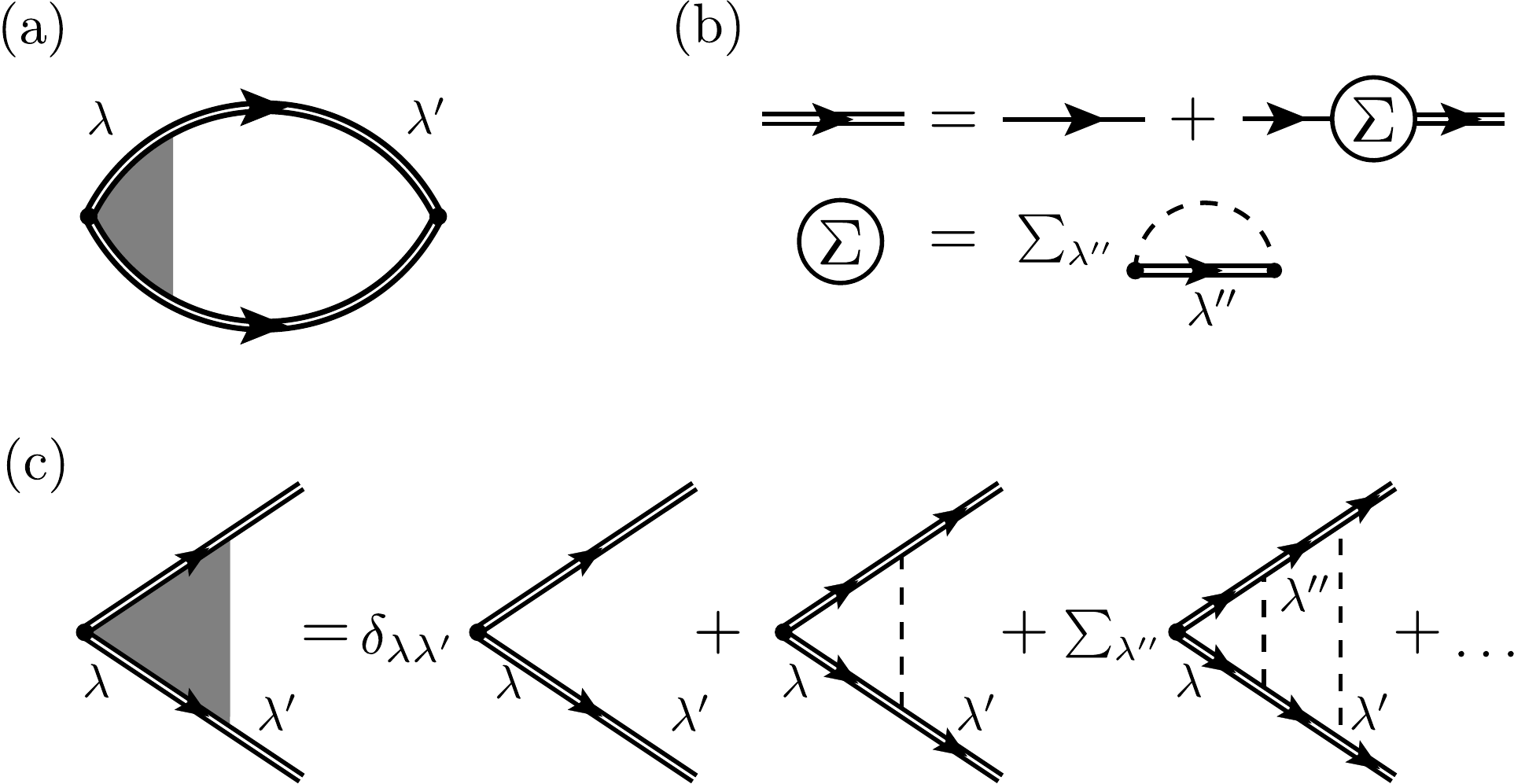}
\caption{As shown in (a), the coefficients $d_{\lambda\lambda'}$ of the Ginzburg-Landau expansion (\ref{AveragedFreeEnergy}) can be represented using the full Greens function (double line) and the dressed vertex (gray triangle). Part (b) and (c) illustrate the diagrams contributing to the Greens function and vertex that are leading in the limit of weak disorder $k_Fl \gg 1$.}
\label{Diagrammatics}
\end{center}
\end{figure}

The disorder averaging in \equref{CoefficientsInExpansion} leads to the effective interaction (\ref{FinalDisorderInteraction}) with the impurity line as defined in \equref{VertexInEigenbasis}. The exact expression for the coefficient $d_{\lambda\lambda'}(T)$ is represented diagrammatically in \figref{Diagrammatics}(a) and requires knowing the full self energy and vertex correction. In this paper, we will assume that the disorder scattering is sufficiently weak in the sense that the mean-free path~$l$ is much larger than the inverse Fermi momentum $k_F^{-1}$. This assumption is justified by the experimental observation that the typical normal state sheet resistance is by more than one order of magnitude smaller than $h/e^2$, where $e$ and $h$ denote the electron charge and Planck's constant.\cite{Reyren,ShalomMFP,ChangMFP} Being suppressed by a factor $(k_F l)^{-1}$, we neglect all diagrams with crossed impurity lines in the self-energy as well as in the vertex correction. Consequently, only the ``rainbow 
diagrams'' and ``Cooperon ladder'', which are shown in \figref{Diagrammatics}(b) and (c), have to be taken into account.

From the diagrammatic representation, one observes that the impurity vertex $\Gamma^{\text{PA}}_{\tau_1 \tau_1',\tau_2 \tau_2'}$ does not enter as a full tensor but only in form of the index combinations
\begin{equation}
\mathcal{S}_{\tau\tau'} =  \Gamma^{\text{PA}}_{\tau \tau',\tau' \tau} = \vcenter{\hbox{\includegraphics[height=3.5em]{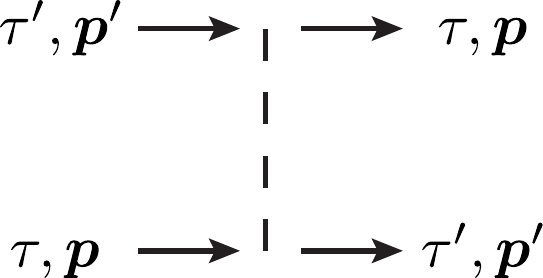}}} \label{SDefinition}
\end{equation} 
and
\begin{align}
 \mathcal{V}_{\tau\tau'} &= \Gamma^{\text{PA}}_{\tau' \tau, \tau'_{\text{K}} \tau^{\phantom{\prime}}_{\text{K}}} = \vcenter{\hbox{\includegraphics[height=3.6em]{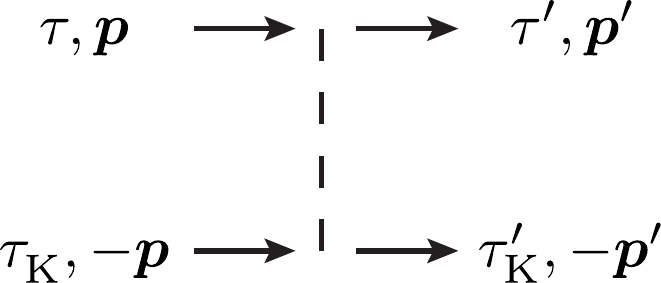}}} \label{VDefinition}
\end{align} 
in the self-energy and vertex correction, respectively. Here the patch momenta $\vec{p}$ and $\vec{p}'$ have been added for later convenience and $\tau_\text{K}=(\lambda,j,-\eta)$ denotes the Kramers partner of patch $\tau=(\lambda,j,\eta)$. In \equref{VDefinition}, we have taken into account that the superconducting vertex only couples Kramers partners, $\{\tau,\vec{p};\tau_\text{K},-\vec{p}\}$, characterized by zero total momentum. As the patches are assumed to be disjoint in momentum space, momentum conservation allows for Cooper pairs only being scattered into pairs of states of the form $\{\tau',\vec{p}';\tau'_\text{K},-\vec{p}'\}$. 

Note that this crucially reduces the complexity of impurity scattering as the tensor $\Gamma^{\text{PA}}$ with $16^4$ possible index combinations is replaced by two scattering matrices $\mathcal{S}$ and $\mathcal{V}$ with only $16^2$ entries each. 
More importantly, we directly see that the momenta $\vec{p}$ and $\vec{p}'$ can freely run over all values within the associated patches since momentum conservation is fulfilled in the diagrams of \equsref{SDefinition}{VDefinition} for arbitrary and independent values of $\vec{p}$ and $\vec{p}'$. We emphasize that this crucial simplification is a consequence of the superconducting state being characterized by Cooper pairs with vanishing center of mass momentum and the absence of diagrams with crossed impurity lines in the diagrammatic expansion of \figref{Diagrammatics}. Had we included diagrams with crossed lines, the impurity vertex would have entered in a more general form with three arbitrary external momenta (one momentum is always fixed by momentum conservation). In this case, also scattering processes of an electron that changes its patch with an electron that stays in the same patch contribute. As discussed in \secref{PatchApprox}, the phase space of these processes strongly depends on quantitative details of the band structure and of the patch geometry. From these considerations, we see that that focusing on the limit $k_Fl \gg 1$ not only allows for a controlled summation of a minimal set of diagrams but also renders the final result independent of the aforementioned microscopic details. 

In the explicit summation and evaluation of the diagrams, which can be found in the \appref, the component of the patch momentum perpendicular to the Fermi surface is constrained by an energetic cutoff $\Lambda$. The longitudinal limits of the patches is absorbed by introducing the density of states $\rho_{\tau}$ per patch. For simplicity, we assume that the density of states does not depend on $\tau$ and introduce $\rho_F= 8\rho_{\tau}$, the density of states per Fermi surface. In \secref{Results}, we will also comment on the modifications in case of an imbalance in the density of states on the different Fermi surfaces.  

Finally, we note that the two scattering matrices in \equsref{SDefinition}{VDefinition} that determine the behavior of the superconductor under the influence of impurity scattering are made unique by the phase convention (\ref{PhaseConventionPatch}):
Obviously, $\mathcal{S}_{\tau\tau'}$ is already invariant under arbitrary $U(1)$-gauge transformations of the wavefunctions, whereas $\mathcal{V}_{\tau\tau'}$ is fixed due to \equref{PhaseConventionPatch} since states with the same $\lambda$ and $j$ but opposite $\eta$ always come in pairs.

\section{Results}
\label{Results}
\begin{figure*}[bt]
\begin{center}
\includegraphics[width=0.8\linewidth]{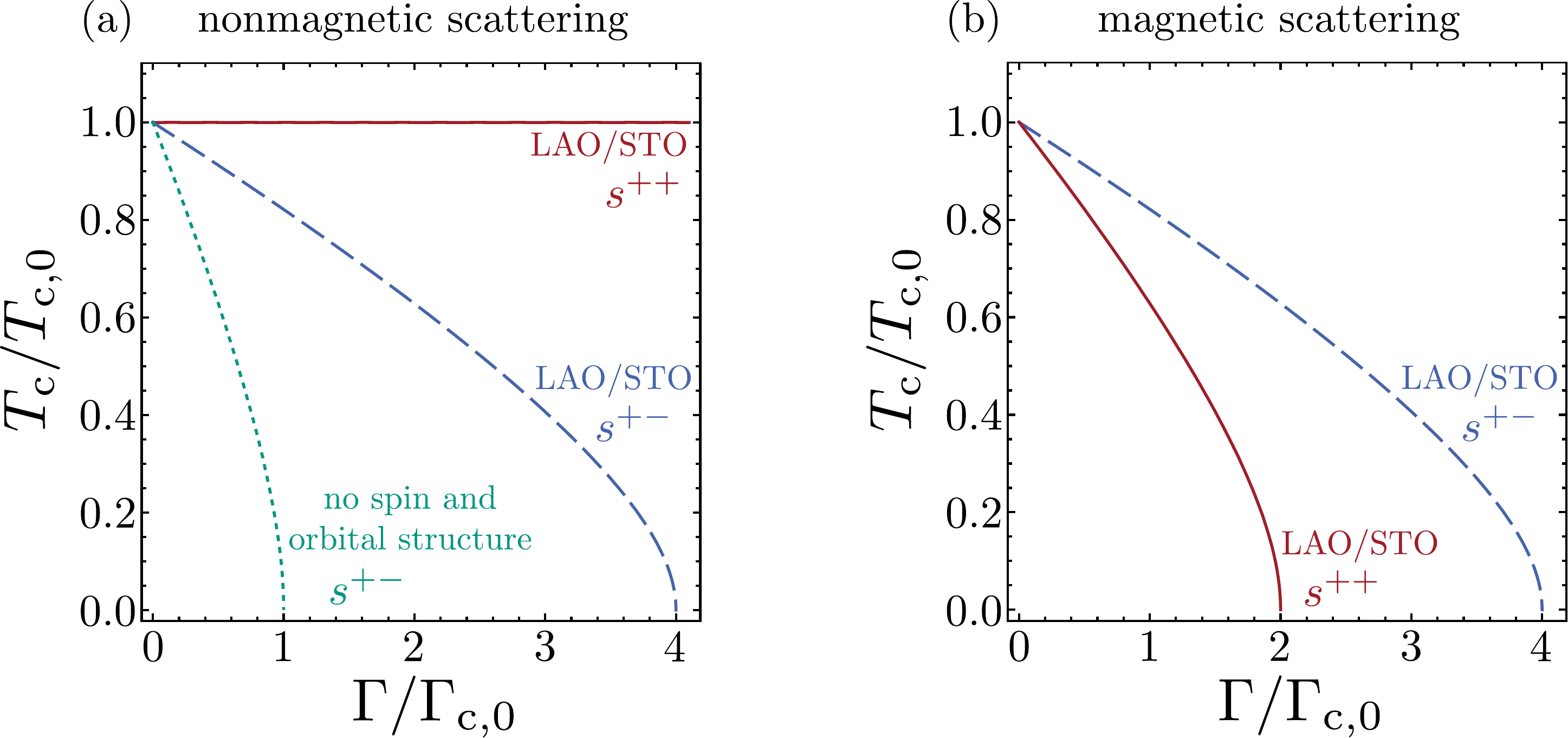}
\caption{The critical temperature $T_c$ as a function of the scattering rate $\Gamma$ for the different superconductors in case of (a) nonmagnetic and (b) magnetic disorder. The scattering rates $\Gamma$ in part (a) and (b) are defined in \equsref{CombinedNonMagnScatteringRate}{CombinedMagnScatteringRate}, respectively. $T_{c,0}$ and $\Gamma_{c,0}$ are the critical temperature (\ref{DefinitionOfCleanTc}) of the clean system and the critical scattering rate (\ref{DefinitionOfTheCriticalGamma}) of the reference model (green dashed line in (a)) defined in the main text.}
\label{ResultsTc}
\end{center}
\end{figure*}

Let us begin our discussion with the clean case. Calculating the Ginzburg-Landau expansion in the absence of any disorder reproduces the results of \refcite{ScheurerSchmalian}: For attractive (repulsive) pair hopping, i.e., $\Vint <0$ ($\Vint >0$) in \equref{PairHoppingInteraction}, the $s^{++}$ ($s^{+-}$) superconductor is realized. In both cases, the transition temperature assumes the usual BCS form
\begin{equation}
 T_{c,0} = \Lambda \frac{2e^\gamma}{\pi}e^{-\frac{2}{\rho_F |\Vint |}}, \label{DefinitionOfCleanTc}
\end{equation} 
with $\gamma$ denoting Euler's constant.

To proceed with the impact of disorder on superconductivity, we first focus on the case of \textit{nonmagnetic disorder} which is characterized by the vertex in \equref{GeneralNonMagn}. 
For attractive pair hopping, we again obtain the $s^{++}$ state. Within the approximations of \secref{GinzburgLandau}, the associated transition temperature is unaffected by disorder irrespective of the values of the different coupling constants in the impurity vertex (\ref{GeneralNonMagn}). This result is a manifestation of the well-known Anderson theorem\cite{AT,ATAG1,ATAG2}.

In case of repulsive pair hopping, $\Vint >0$, we still find that, as in the clean case, the $s^{+-}$ superconductor condenses. In accordance with the usual expectation that unconventional superconductors are less robust against impurity scattering, the transition temperature is now reduced by disorder. 
Despite the complicated form of the vertex (\ref{GeneralNonMagn}), it turns out that the impurity distribution only enters the Ginzburg-Landau expansion and, thus, the transition temperature, in form of the combined scattering rate
\begin{equation}
 \Gamma = \gamma_{A_1}^{\text{I}} + \gamma_{A_1}^{\text{II}} + \gamma_{B_1} + \gamma_{B_2} + \gamma_E. \label{CombinedNonMagnScatteringRate}
\end{equation} 
Note that $\gamma_{A_1}^{\text{III}}$, which describes the admixture of $s$-wave and spin-orbit disorder on the same site, does not enter at all. 
This follows from the strong orbital polarization of the wavefunctions in the patches (see \figref{SpinTexturesOfTheModel}), which we have idealized to be perfect in our calculation. It is easily seen upon inserting the term with $\gamma_{A_1}^{\text{III}}$ in \equref{GeneralNonMagn} into the two relevant scattering matrices defined in \equsref{SDefinition}{VDefinition}. Since the two scattered electrons either both switch the orbital or both keep their orbital character, either $\tau_0\sigma_0$ or $\tau_2\sigma_3$ in \equref{GeneralNonMagn} yields zero in $\mathcal{S}$ and $\mathcal{V}$.
Furthermore, we emphasize that the effective scattering strength and, hence, the transition temperature do not depend on the microscopic details of the impurity scattering, i.e., on the relative weight between different terms in \equref{GeneralNonMagn}. As the different coupling constants are proportional to the concentration of the impurities and the square of the matrix element of the associated local perturbations, $\Gamma$ in \equref{CombinedNonMagnScatteringRate} measures the average overall strength of the disorder configuration.

The transition temperature $T_c$ is found from demanding that the corresponding eigenvalue in the disorder averaged Ginzburg-Landau expansion (\ref{AveragedFreeEnergy}) changes sign. This leads to (see \appref)
\begin{equation}
 0 = \frac{1}{|\Vint |} - \frac{\rho_F}{2} \left[ \ln\left(\frac{\Lambda}{2\pi T_c}\right) - \psi_0 \left(\frac{1}{2}+e^{-\gamma}\frac{\Gamma/\Gamma_c}{T_c/T_{c,0}} \right) \right], \label{EigenvalueEquationTc}
\end{equation} 
where $\psi_0$ denotes the digamma function and $\Gamma_c = e^{-\gamma}T_{c,0}/\rho_F$ has been introduced for convenience.
\equref{EigenvalueEquationTc} yields the functional dependence $T_c/T_{c,0} = f(\Gamma/\Gamma_c)$ of the transition temperature on the scattering strength $\Gamma$. Mathematically, it reproduces the usual Abrikosov-Gorkov law \cite{AGLaw} originally describing the critical temperature of BCS superconductors under the influence of spin-magnetic disorder. Analytically, one can easily show that $f(1) = 0$, which identifies $\Gamma_c$ as the critical scattering rate, as well as
\begin{equation}
  f(\Gamma/\Gamma_c) \sim 1 - \frac{\pi^2}{2 e^\gamma} \frac{\Gamma}{\Gamma_c}
\end{equation} 
for $\Gamma/\Gamma_c \rightarrow 0$. The functional behavior for general values of $\Gamma \leq \Gamma_c$ is shown in \figref{ResultsTc}(a) (together with the Anderson theorem for $s_{++}$ discussed above). 
In this figure, the scattering strength is measured in units of the reference critical scattering rate $\Gamma_{c,0}$. This quantity is defined by considering the situation where the patch wavefunctions would allow electrons to scatter between all patches in a way that does not break time-reversal symmetry (nonmagnetic scattering). It corresponds to fermions with the same band structure but without spin and orbital degrees of freedom. Then $\Gamma_{c,0}$ is the critical value of $\Gamma = \mathcal{S}_{\tau,\tau'}$ at which the $s^{+-}$ state is fully destroyed. To be explicit, one finds
\begin{equation}
 \Gamma_{c,0} = \frac{T_{c,0}}{4e^{\gamma}\rho_F}. \label{DefinitionOfTheCriticalGamma}
\end{equation} 
The enhancement $\Gamma_c = 4 \Gamma_{c,0}$ of the critical scattering rate of $s^{+-}$ as compared to the reference model is due to the locking between the crystal momentum and the spin/orbital degree of freedom. To illustrate this point, consider the simplest case of spin-trivial $s$-wave impurities, \ie only take the first term in \equref{GeneralNonMagn}. Recalling the spin and orbital textures of \figref{SpinTexturesOfTheModel}, an electron from, e.g., patch $(1,2,+)$ in \figref{IllustrationOfPatchNotation} cannot scatter into one of the patches of the perpendicular lobes of the Fermi surfaces since it cannot change its orbital character. Within the same lobe, it can only scatter into states of the patches $(1,1,+)$ and $(2,2,-)$ as a consequence of the spin polarization of the wavefunctions. We thus see that the spin and orbital polarization of the Fermi surfaces reduces the possibilities for scattering events as compared to naively neglecting the spin and orbital degrees of freedom of the electrons.
Note that this mechanism of protection against impurity scattering can in some systems even rule out all scattering processes that are pair breaking as has been demonstrated in \refcite{SpinOrbitLocking}.

By independently rescaling the intra- and interband scattering rate (see \appref), we have verified that only the interband processes affect the transition temperature of the $s^{+-}$ superconductor. 
This is expected as, in the absence of interband scattering, the $s^{++}$ and $s^{+-}$ superconductors have to behave identically and, hence, both be unaffected by disorder due to the Anderson theorem.

Let us next discuss \textit{magnetic disorder} which is, in its most general form, defined by the impurity vertex in \equref{MagneticImp}. Similarly to the nonmagnetic case, the impurity distribution only enters in form of the single variable 
\begin{equation}
 \Gamma = \gamma_{E}^{\text{I}} + \gamma_{E}^{\text{II}} + \gamma_{E}^{\text{III}} + \gamma_{A_2}^{\text{I}} + \gamma_{A_2}^{\text{II}} + \gamma_{B_1} + \gamma_{B_2} \label{CombinedMagnScatteringRate}
\end{equation}
in the expansion (\ref{AveragedFreeEnergy}) of the average free energy. This shows that also the impact of magnetic scattering on superconductivity does not depend on microscopic details of the impurities involved. It is even independent of whether the magnetic moments are aligned in-plane (first three terms in \equref{CombinedMagnScatteringRate}) or perpendicular to it (remaining four coupling constants).
Note that $\gamma_{A_2}^{\text{III}}$ in \equref{MagneticImpPerp} does not contribute at all, which is, exactly as in case of $\gamma_{A_1}^{\text{III}}$ above, due to the spin and orbital polarization of the patch wavefunctions.

We find that, as before, $\Vint <0$ ($\Vint >0$) leads to pure $s^{++}$ ($s^{+-}$) superconductivity. However, in case of magnetic impurities, both superconductors are affected by disorder with the transition temperature again following from \equref{EigenvalueEquationTc} where $\Gamma_c = 2\Gamma_{c,0}$ and $\Gamma_c = 4\Gamma_{c,0}$ in case of $s^{++}$ and $s^{+-}$, respectively. The resulting dependence $T_c/T_{c,0} = f(\Gamma/\Gamma_c)$ of the transition temperature on the scattering strength $\Gamma$ in \equref{CombinedMagnScatteringRate} is shown in \figref{ResultsTc}(b).

Interestingly enough, the unconventional $s^{+-}$ superconductor is (by a factor of $2$) more robust against magnetic scattering than the conventional $s^{++}$ state. To understand this, we have again analyzed which type of scattering processes act pair breaking in the two superconductors. One finds that $s^{++}$ is suppressed both by inter- and intraband scattering, whereas the $s^{+-}$ state is only sensitive to intraband processes. This leads to the relative factor of $2$ mentioned above. 
We refer to \refcite{PnictidesScattering} for another example of the stability against TRA interband scattering of a superconductor with an order parameter that changes sign between two Fermi surfaces.
The additional factor of $2$ with respect to $\Gamma_{c,0}$ is, again, due to reduction of possible scattering processes resulting from the spin and orbital polarization of the patch wavefunctions.

Finally, a few remarks are in order that refer both to magnetic and nonmagnetic scattering.
Firstly, taking the spin and orbital polarization of the patch wavefunctions to be perfect constitutes a very good approximation as can be seen in \figref{SpinTexturesOfTheModel}. Introducing deviations from perfect polarization would have the following major modifications: The impact of disorder on the superconductors would become less universal in the sense that it would not be just the sum of all coupling constants as in \equsref{CombinedNonMagnScatteringRate}{CombinedMagnScatteringRate} determining the transition temperature. Furthermore, the enhancement $\Gamma_c/\Gamma_{c,0}$ of the critical scattering rates with respect to the reference model would be reduced as the orbital/spin-momentum locking would be weaker.

Secondly, having slightly different densities of states on the two Fermi surfaces would not crucially alter our conclusions. Again, the results would become less universal, since, already in the clean system, we would get an admixture of $s^{++}$ and $s^{+-}$ superconductivity. Then, e.g., for $\Vint <0$ and nonmagnetic disorder, the order parameter of the resulting superconductor would have identical signs on both Fermi surfaces but slightly different magnitudes. We would find that this superconductor is destroyed by disorder at some finite but very large critical scattering strength.

Furthermore, we emphasize that we have discussed the most general model of local, $\delta$-correlated nonmagnetic and magnetic disorder here. Any more specific model is thus covered by \equsref{GeneralNonMagn}{MagneticImp}. One might be interested, e.g., in the case where not only the disorder distribution but also every single impurity has the full symmetry of the system (only $\gamma_{A_1}^{\text{I}}$, $\gamma_{A_1}^{\text{II}}$ and $\gamma_{A_1}^{\text{III}}$ in \equref{GeneralNonMagn} are finite). To give another example, isotropic in-plane spin-magnetism corresponds to $\gamma_{E}^{\text{I}}$ in \equref{MagneticImpPara}. 
\equsref{CombinedNonMagnScatteringRate}{CombinedMagnScatteringRate} show that, despite the plethora of possible microscopic scenarios in the two-band model of oxide interfaces, only the four different combinations $\Vint >0$/$\Vint <0$ and magnetic/nonmagnetic disorder need to be discussed separately.

\section{Conclusion}
To summarize, we have considered a model for the itinerant electrons at oxide interfaces that includes the Ti $3d_{xz}$ and $3d_{yz}$ orbitals which are expected to be most relevant for describing superconductivity\cite{PentchevaPickett1,PentchevaPickett2,PotterLee,SpinSplitting,Joshua2012,TrisconeConfExp}. Depending on microscopic details of the electron-electron interaction, this model yields\cite{ScheurerSchmalian} a conventional $s^{++}$ or an unconventional $s^{+-}$ superconductor.

To describe disorder scattering, we have taken the most general $\delta$-correlated impurity distribution that is consistent with the spatial symmetries of the system (point group $C_{4v}$ and 2D translation) as well as with time-reversal symmetry. Note that these symmetries only have to be restored on average such that a single impurity can locally break any symmetry of the system. The combination of spin and orbital degrees of freedom leads to many different possible scattering channels as can be seen in the impurity vertices in \equsref{GeneralNonMagn}{MagneticImp} for nonmagnetic (TRS) and magnetic (TRA) impurities, respectively. 
We have discussed how these different processes can emerge due to vacancies or the formation of local magnetic moments.

As very different behavior is expected, we have studied the impact of magnetic and nonmagnetic disorder on the two superconductors $s^{++}$ and $s^{+-}$ separately. In all cases, we find that, below the critical scattering rates, the form of the superconducting order parameter is unaffected by disorder. Despite the multitude of possible scattering channels, the information of the impurity statistics only enters in form of the combined scattering rates $\Gamma$ defined in \equsref{CombinedNonMagnScatteringRate}{CombinedMagnScatteringRate} for nonmagnetic and magnetic scattering, respectively. 
Our results for the transition temperature of the disordered superconductors, which are summarized in \figref{ResultsTc}, lead to the following two central conclusions: 
Firstly, the critical value of the scattering rate $\Gamma$, where both superconductors in case of magnetic and $s^{+-}$ in case of nonmagnetic impurities are fully destroyed, is enhanced as compared to a reference model that naively neglects the spin and orbital polarization. Although just being a factor of order $1$, this enhancement of stability against disorder might be relevant for oxide interfaces where the mean-free path\cite{ShalomMFP,ChangMFP} is expected to be comparable to the coherence length\cite{Reyren}. 
Secondly, contrary to the expectation that unconventional superconductivity is less robust in the presence of disorder, we find that the critical scattering rate of $s^{+-}$ is twice as large as that of the $s^{++}$ superconductor in case of magnetic disorder. We emphasize that magnetic impurities are expected to be particularly relevant in oxide interfaces as the system has the tendency towards the formation of a magnetic instability\cite{PentchevaPickett1,PentchevaPickett2,PotterLee,SpinSplitting}. For example, oxygen vacancies have been shown\cite{DFToxygen1,DFToxygen2} to induce local magnetic moments on the Ti sites.
Taken together, these findings indicate that the presence of superconductivity despite the coherence length and mean-free path being comparable does not allow for ruling out unconventional pairing. 

\appendix*
\section{Summation of the diagrams of the Ginzburg-Landau expansion}
\label{SummationOfDiagrams}

This appendix provides details on the calculation of the coefficients $d_{\lambda\lambda^\prime}(T)$ in the disorder-averaged expansion of the free energy in terms of the order parameters as defined in Eqs.~\eqref{AveragedFreeEnergy} and \eqref{CoefficientsInExpansion}, and diagrammatically represented in Fig.~\ref{Diagrammatics}. The approach we present is tailored to the description of oxide heterostructures in the patch approximation (Sec.~\ref{PatchApprox}) of the Fermi surfaces. 
However, the described procedure can be modified straightforwardly for application to other multiband systems as long as the Fermi surfaces can be partitioned in a similar way that allows for a formulation of the problem in terms of scattering rates that are constant within each patch. 

In the presence of disorder, the full propagator (cf. Fig.~\ref{Diagrammatics}(b)) in patch~$\tau$ acquires a finite lifetime. It is modified by a finite self energy~$\Sigma_\tau(\vec{p},\omega_n)$,
\begin{equation}
 G_{\tau}(\vec{p},\omega_n)=\frac{1}{i\omega_n-\epsilon_{\tau}(\vec{p})-\Sigma_{\tau}(\vec{p},\omega_n)}, 
\end{equation}
where $\epsilon_\tau(\vec{p})$ denotes the disperson in patch~$\tau$. 
We consider weak disorder such that diagrams with crossed impurity lines are suppressed by a factor $(k_Fl)^{-1}$ and calculating the self energy in Born approximation is sufficient. 
These assumptions result in the expression 
\begin{equation}
 \Sigma_\tau(\vec{p},\omega_n)=-i\pi \sign (\omega_n)\sum_{\tau^\prime}\rho_{\tau^\prime}\mathcal{S}_{\tau\tau^\prime}
\end{equation}
for the self energy in patch~$\tau$, where $\rho_\tau$ is the density of states in patch~$\tau$, and the corresponding scattering matrix~$\mathcal{S}_{\tau\tau^\prime}$ of scattering processes contributing to the self energy in the patch approximation is defined in Eq.~\eqref{SDefinition}. As introduced in  Sec.~\ref{GinzburgLandau}, we assume that the density of states does not depend on~$\tau$ and use $\rho_F=8\rho_\tau$ in the remainder of this appendix.   

\vspace{0.9em}

The expansion coefficients $d_{\lambda\lambda^\prime}(T)$ in the presence of disorder contain contributions of all possible noncrossing combinations of the allowed scattering processes of Kramers partners (cf. the definition of $\mathcal{V}_{\tau\tau^\prime}$ in Eq.~\eqref{VDefinition}), 
\begin{align}
\begin{split}
 \vcenter{\hbox{\includegraphics[height=3.0em]{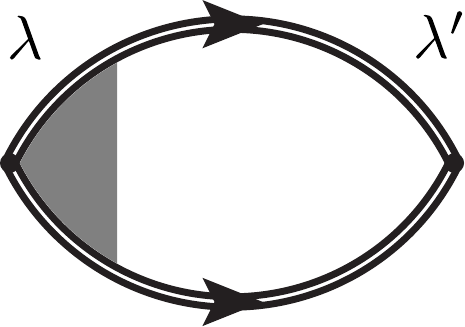}}} \, \, &= \,\, \delta_{\lambda\lambda^\prime}\vcenter{\hbox{\includegraphics[height=3.0em]{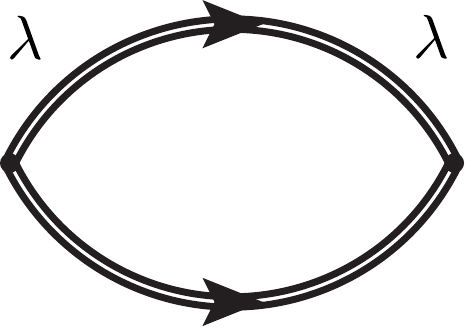}}}\,\, + \,\,\vcenter{\hbox{\includegraphics[height=3.0em]{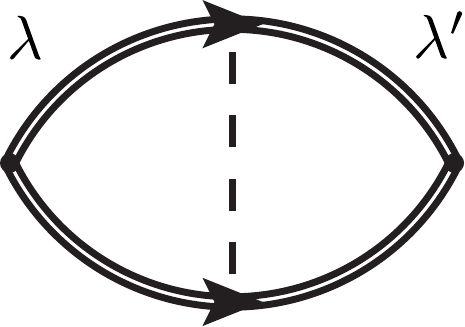}}} \\ &+\sum_{\lambda^{\prime\prime}}\vcenter{\hbox{\includegraphics[height=3.0em]{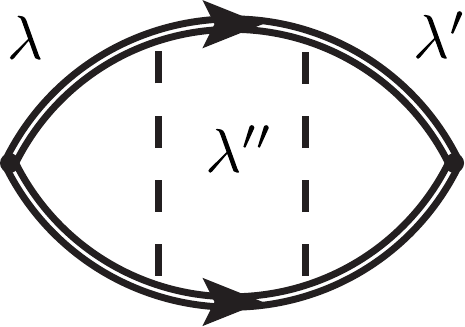}}} \,\, + \, \sum_{\lambda_1,\lambda_2}\vcenter{\hbox{\includegraphics[height=3.0em]{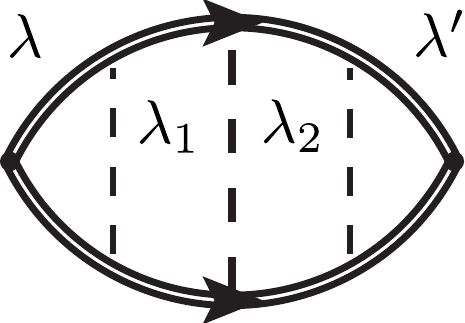}}}\,\, +\, \,\ldots\label{CoefficientDiagrammaticSum}\end{split}
\end{align}

For within the patch approximation, the impurity vertex is a constant $\mathcal{V}_{\tau\tau^\prime}$ for all scattering processes of Kramers partners from patch~$\tau$ to patch~$\tau^\prime$, the momentum integrals in Eq.~\eqref{CoefficientDiagrammaticSum} decouple and it is useful to define 
\begin{equation}
  c_{\tau}(\omega_n)=\sum_{\vec{p}}G_{\tau}(\vec{p},\omega_n)G_{\tau_{\text{K}}}(-\vec{p},-\omega_n) 
\end{equation}
which evaluates to 
 \begin{equation}
  c_{\tau}(\omega_n)= \frac{\pi\rho_F}{8|\omega_n|+\frac{\pi\rho_F}{2}\sign(\omega_n)\sum_{\tau'}[\mathcal{S}_{\tau\tau'}+\mathcal{S}_{\tau_{\text{K}}\tau'}]},
 \end{equation}
where we used $\epsilon_{\tau_{\text{K}}}(-\vec{p})=\epsilon_\tau(\vec{p})$. 

Exploiting the patch approximation and following the notation introduced in Sec.~\ref{GinzburgLandau}, the coefficients in the Ginzburg-Landau expansion can be written in the form 
\begin{widetext}
\begin{align}
 d_{\lambda\lambda^\prime}&=-\delta_{\lambda\lambda^\prime}\frac{T}{2}\sum_{\omega_n}\sum_{j,\eta}c_{(\lambda,j,\eta)}(\omega_n)-\frac{T}{2}\sum_{\omega_n}\sum_{j,j^\prime}\sum_{\eta,\eta^\prime}\eta\eta^\prime c_{(\lambda,j,\eta)}(\omega_n)\mathcal{V}_{(\lambda,j,\eta)(\lambda^\prime,j^\prime,\eta^\prime)}c_{(\lambda^\prime,j^\prime,\eta^\prime)}(\omega_n) 
 \nonumber \\ 
 & \qquad - \frac{T}{2}\sum_{\omega_n}\sum_{\lambda^{\prime\prime}}\sum_{j,j^\prime,j^{\prime\prime}}\sum_{\eta,\eta^\prime,\eta^{\prime\prime}}\eta\eta^\prime c_{(\lambda,j,\eta)}(\omega_n)\mathcal{V}_{(\lambda,j,\eta)(\lambda^{\prime\prime},j^{\prime\prime},\eta^{\prime\prime})}c_{(\lambda^{\prime\prime},j^{\prime\prime},\eta^{\prime\prime})}(\omega_n)\mathcal{V}_{(\lambda^{\prime\prime},j^{\prime\prime},\eta^{\prime\prime})(\lambda^\prime,j^\prime,\eta^\prime)}c_{(\lambda^\prime,j^\prime,\eta^\prime)}(\omega_n)+\ldots,\label{CoefficientSum}
\end{align}
\end{widetext}
suggesting to introduce, analogously to $\mathcal{S}$ and $\mathcal{V}$, nested diagonal matrices $\mathcal{C}$ containing the propagator products $c_{\tau}(\omega_n)$,
\begin{equation}
  \mathcal{C}_{\tau\tau'} = \delta_{\tau\tau^\prime}c_{\tau}(\omega_n).
\end{equation}
With these definitions at hand, we can conveniently express Eq.~\eqref{CoefficientSum} in matrix form, 
\begin{equation}
 d_{\lambda\lambda^\prime}=-\frac{T}{2}\sum_{\omega_n}\sum_{j,j^\prime}\sum_{\eta,\eta^\prime}\eta\eta^\prime\left[\sum_{m=0}^\infty\left(\mathcal{C}\mathcal{V}\right)^m	\mathcal{C}\right]_{(\lambda,j,\eta)(\lambda^\prime,j^\prime,\eta^\prime)} %
\end{equation}
and therefore, the results for the coefficients $d_{\lambda\lambda^\prime}(T)$ can be obtained from matrix multiplication and matrix inversion, 
\begin{equation}
d_{\lambda\lambda^\prime}= -\frac{T}{2}\sum_{\omega_n}\sum_{j,j^\prime}\sum_{\eta,\eta^\prime}\eta\eta^\prime\left[\left(\mathds{1}-\mathcal{C}\mathcal{V}\right)^{-1}\mathcal{C}\right]_{(\lambda,j,\eta)(\lambda^\prime,j^\prime,\eta^\prime)}.
\end{equation}
This result can be staightforwardly applied to other multiband superconductors, if an analogous partitioning of the Fermi surface allows to formulate the problem in terms of scattering rates that are only characterized by respective patch indices and, apart from that, do not depend on the momenta of the scattered states.

We investigate the different roles of intraband and interband scattering processes for nonmagnetic and magnetic disorder by artificially discriminating intraband ($\lambda=\lambda^\prime$) and interband ($\lambda\neq\lambda^\prime$) scattering processes by setting 
\begin{align}
 \mathcal{S}_{(\lambda,j.\eta)(\lambda^\prime,j^\prime,\eta^\prime)}&\rightarrow\left\{\begin{matrix} \chi_\mathrm{intra}\mathcal{S}_{(\lambda,j,\eta)(\lambda^\prime,j^\prime,\eta^\prime)}, & \lambda=\lambda^\prime, \\ \chi_\mathrm{inter} \mathcal{S}_{(\lambda,j,\eta)(\lambda^\prime,j^\prime,\eta^\prime)}, & \lambda\neq\lambda^\prime,\end{matrix}\right. \\ 
 \mathcal{V}_{(\lambda,j.\eta)(\lambda^\prime,j^\prime,\eta^\prime)}&\rightarrow\left\{\begin{matrix} \chi_\mathrm{intra}\mathcal{V}_{(\lambda,j,\eta)(\lambda^\prime,j^\prime,\eta^\prime)}, & \lambda=\lambda^\prime, \\ \chi_\mathrm{inter} \mathcal{V}_{(\lambda,j,\eta)(\lambda^\prime,j^\prime,\eta^\prime)}, & \lambda\neq\lambda^\prime.\end{matrix}\right.
\end{align}
In order to reproduce \equref{EigenvalueEquationTc} and the dependence of the transition temperature~$T_c$ on the scattering rate~$\Gamma$ as shown in Fig.~\ref{ResultsTc}, the results discussed in the remainder of this appendix have of course to be understood in the sense that $\chi_\mathrm{intra}=\chi_\mathrm{inter}=1$. 

The resulting diagrams in the patch approximation satisfy $d_{11}(T)=d_{22}(T)$ and $d_{12}(T)=d_{21}(T)$, as well as $d_{11}\in\mathbb{R}$ and $d_{12}\in\mathbb{R}$, and thereby, the free energy~\eqref{AveragedFreeEnergy} can be diagonalized as 
\begin{align}
  \braket{\mathcal{F}}_\text{dis} &\sim \Big[\frac{-1}{\Vint }+d_{11}(T)+d_{12}(T)\Big]\left|\tfrac{1}{\sqrt{2}}\left(\Delta_1+\Delta_2\right)\right|^2
   \nonumber \\ 
  &  \quad + \Big[\frac{1}{\Vint }+d_{11}(T)-d_{12}(T)\Big]\left|\tfrac{1}{\sqrt{2}}\left(\Delta_1-\Delta_2\right)\right|^2. 
\end{align}
The resulting transition temperature for either $s^{++}$ ($\Delta_1=\Delta_2$, in case of $\Vint <0$) or $s^{+-}$ ($\Delta_1=-\Delta_2$, in case of $\Vint >0$) superconductivity is found from demanding that the corresponding eigenvalue change sign at~$T=T_c$.
\paragraph{Nonmagnetic disorder}
In case of disorder potentials that are symmetric under time-reversal (TRS), that is, for nonmagnetic disorder, we find 
\begin{align}
 d_{11}(T)&=-\frac{\rho_F}{4}\Big[2\ln\Big(\frac{\Lambda}{2\pi T}\Big)  -\operatorname{\psi_0}\Big(\frac{1}{2}\Big)  \nonumber \\ 
  &\qquad -\operatorname{\psi_0}\Big(\frac{1}{2}+\frac{\rho_F\chi_\mathrm{inter}\Gamma}{4T}\Big)\Big],\\
 d_{12}(T)&=-\frac{\rho_F}{4}\Big[ 
 -\operatorname{\psi_0}\Big(\frac{1}{2}\Big) \nonumber \\ &\qquad +\operatorname{\psi_0}\Big(\frac{1}{2}+\frac{\rho_F\chi_\mathrm{inter}\Gamma}{4 T}\Big)\Big], 
\end{align}
and consequently, the transition temperature for the $s^{++}$ superconducting state in the presence of nonmagnetic impurities is given by the condition 
\begin{equation}
  0=\frac{1}{|\Vint |}-\frac{\rho_F}{2}\Big[\ln\Big(\frac{\Lambda}{2\pi T}\Big)-\operatorname{\psi_0}\Big(\frac{1}{2}\Big)\Big],
\end{equation}
resulting in a constant transition temperature that coincides with the transition temperature in the clean case, 
\begin{equation}
 T_{c}=\Lambda\frac{2 e^\gamma}{\pi}e^{-\frac{2}{\rho_F|\Vint |}}=T_{c,0}, 
\end{equation}
that is, we reproduce the Anderson theorem\cite{AT,ATAG1,ATAG2}. 
The corresponding condition for $s^{+-}$ superconductivity in the presence of nonmagnetic disorder is
\begin{equation}
  0=\frac{1}{|\Vint |}-\frac{\rho_F}{2}\Big[\ln\Big(\frac{\Lambda}{2\pi T}\Big)-\operatorname{\psi_0}\Big(\frac{1}{2}+\frac{\rho_F\chi_\mathrm{inter}\Gamma}{4T}\Big)\Big], 
\end{equation}
meaning that only interband scattering processes affect the transition temperature of the $s^{+-}$~pairing state, and the suppression of the transition temperature $T_c$ with increasing scattering rate~$\Gamma$ is described by an Abrikosov-Gorkov law\cite{AGLaw} with critical scattering rate $\Gamma_c=e^{-\gamma}T_{c,0}/\rho_F$. 
\paragraph{Magnetic disorder} In the presence of time-reversal antisymmetric (TRA), \ie magnetic, disorder, we find for the expansion coefficients 
\begin{align}
 d_{11}(T)&=-\frac{\rho_F}{4}\Big[2\ln\Big(\frac{\Lambda}{2\pi T}\Big) 
 -\operatorname{\psi_0}\Big(\frac{1}{2}+\frac{\rho_F\chi_\mathrm{intra}\Gamma}{4T}\Big) \nonumber\\ 
  & \qquad -\operatorname{\psi_0}\Big(\frac{1}{2}+\frac{\rho_F(\chi_\mathrm{intra}+\chi_\mathrm{inter})\Gamma}{4T}\Big)\Big],\\
 d_{12}(T)&=-\frac{\rho_F}{4}\Big[\operatorname{\psi_0}\Big(\frac{1}{2}+\frac{\rho_F\chi_\mathrm{intra}\Gamma}{4 T}\Big) \nonumber \\ 
 &\qquad -\operatorname{\psi_0}\Big(\frac{1}{2}+\frac{\rho_F(\chi_\mathrm{intra}+\chi_\mathrm{inter})\Gamma}{4T}\Big)\Big].
\end{align}
The transition temperature of the $s^{++}$~pairing state is thus determined by the condition
\begin{equation}
  0=\frac{1}{|\Vint |}-\frac{\rho_F}{2}\Big[\ln\Big(\frac{\Lambda}{2\pi T}\Big)-\operatorname{\psi_0}\Big(\frac{1}{2}+\frac{\rho_F(\chi_\mathrm{intra}+\chi_\mathrm{inter})\Gamma}{4T}\Big)\Big],
\end{equation}
yielding an Abrikosov-Gorkov-type suppression with a critical scattering rate of $\Gamma_c=e^{-\gamma}T_{c,0}/(2\rho_F)$ since $s^{++}$~superconductivity in the presence of magnetic disorder is destroyed by intraband as well as by interband scattering processes. In contrast, $s^{+-}$~superconductivity is only affected by intraband scattering processes, as seen from the corresponding condition 
\begin{equation}
  0=\frac{1}{|\Vint |}-\frac{\rho_F}{2}\Big[\ln\Big(\frac{\Lambda}{2\pi T}\Big)-\operatorname{\psi_0}\Big(\frac{1}{2}+\frac{\rho_F\chi_\mathrm{intra}\Gamma}{4T}\Big)\Big], 
\end{equation}
and thus the respective critical scattering rate is again given by $\Gamma_c=e^{-\gamma}T_{c,0}/\rho_F$. 

%

\end{document}